\documentclass[10pt]{article}
\usepackage{authblk}
\usepackage{anysize,graphicx}
\usepackage{hyperref}
\usepackage{pstricks,amsmath,amsthm,amsfonts}
\usepackage{pst-node}
\usepackage{pst-tree}
\usepackage{caption}

\title{DSA-aware multiple patterning for the manufacturing of vias: Connections to graph coloring problems, IP formulations, and numerical experiments}
\author[3]{Dehia Ait-Ferhat}
\author[1]{Vincent Juliard}
\author[2]{Gautier Stauffer}
\author[1]{Juan Andres Torres}
\affil[1]{Mentor Graphics Corporation. E-mail: \{vincent\_juliard@mentor.com, andres\_torres@mentor.com}
\affil[2]{Center of Excellence in Supply Chain Innovation and Transportation (CESIT). Kedge Business School, Talence, France. E-mail: gautier.stauffer@kedgebs.com}
\affil[3]{dehia.aitferhat@gmail.com}
\date{}

\begin{document}
\maketitle

\abstract{In this paper, we investigate the manufacturing of vias in integrated
circuits with a new technology combining
lithography and Directed Self Assembly (DSA). Optimizing the
production time and costs in this new process entails minimizing the
number of lithography steps, which constitutes a generalization of graph
coloring. We develop integer programming formulations for
several variants of interest in the industry, and then study the
computational performance of our formulations on true industrial
instances. We show that the best integer programming formulation
achieves good computational performance, and indicate potential
directions to further speed-up computational time and develop exact
approaches feasible for production.
 }

\section{Preliminaries}\label{introduction}

For the past decades, one of the main drivers of the explosion in the adoption of electronic components in our daily lives has been the addition of more functionality at a lower cost. This has traditionally been achieved by scaling down the geometries in the devices. At every {\em technology node}\footnote{A technology node refers to a semiconductor manufacturing process. It usually takes the form of a distance, e.g., 193nm, 22nm, and is historically closely linked to chip density. \url{https://en.wikichip.org/wiki/technology_node}}, new production methods allow devices to occupy less total space while at the same time enabling other properties, such as lower power consumption and faster switching.

However, in the last few years, the challenge of continuing on this rapid trajectory to ever-smaller feature sizes has increased: moving from 193nm lasers to a 13nm wavelength (in Extreme Ultra-Violet (EUV)) has required the complete redesign of lithography systems from optical diffraction to reflection projection systems. At 13nm, this radiation is mostly absorbed in the materials, rather than diffracted or reflected. Due to all the challenges associated with this technology, the industry has started using multiple patterning techniques where the design is separated into multiple patterning steps when a dense pattern in a single exposure is not possible. At and below 22nm technology nodes, it is impossible to reproduce the intended features using a single lithographic step, and the industry has thus resorted to using double, and in some cases triple, patterning.

However, the move to multiple patterning also has scaling implications. It reduces the total throughput of the system, and while a piece of equipment could previously process {\em N} number of wafers, now the actual number is {\em N/i} where {\em i} is the number of patterning steps. So while denser patterns can be achieved, the process does not become immediately more cost-effective.

As a consequence, interest has grown in finding process technologies that cost-effectively reduce the total number of patterning steps. While EUV is one such technique, the investment in new lithography equipment, 13nm light sources, power requirements, and the development of new production materials has led to the search for alternatives. DSA (Directed Self Assembly), is one of these techniques that can in principle achieve finer feature sizes with a lower number of patterning steps.

The self-assembly process uses the thermodynamic properties of diblock copolymers to form lines or circles on a surface \cite{DSAdepablo}.
These structures are randomly formed and controlled by the diblock copolymer architecture. The main idea is to chemically join two different types of polymers, such as polystyrene (PS) and poly(methyl methacrylate) (PMMA). Unless chemically bonded, they would separate at a macro level. However, when a molecule is composed of half PS and half PMMA, the molecules cannot macro separate, and therefore align in ways where PS attempts to surround itself with other PS segments, and PMMA with other PMMA segments.

As random micro-patterns are not very useful in semiconductor manufacturing, guiding patterns can be shaped that {\em direct} how the material alignment will take place. The idea is to exploit the diblock copolymer’s properties to achieve the necessary assembly to transfer the desired pattern onto a wafer. Ingeniously combining adequate guiding patterns with multiple patterning can then help reduce the number of patterning steps. This process is referred to as DSA-aware multiple patterning. In this paper, we investigate the corresponding process from an optimization point of view, starting with a gentle introduction for non-experts.

\section{A gentle introduction to DSA-aware multiple patterning}  

\begin{figure}[h!]
	\begin{center}
	\begin{tabular}{cc}\includegraphics[scale = 0.27]{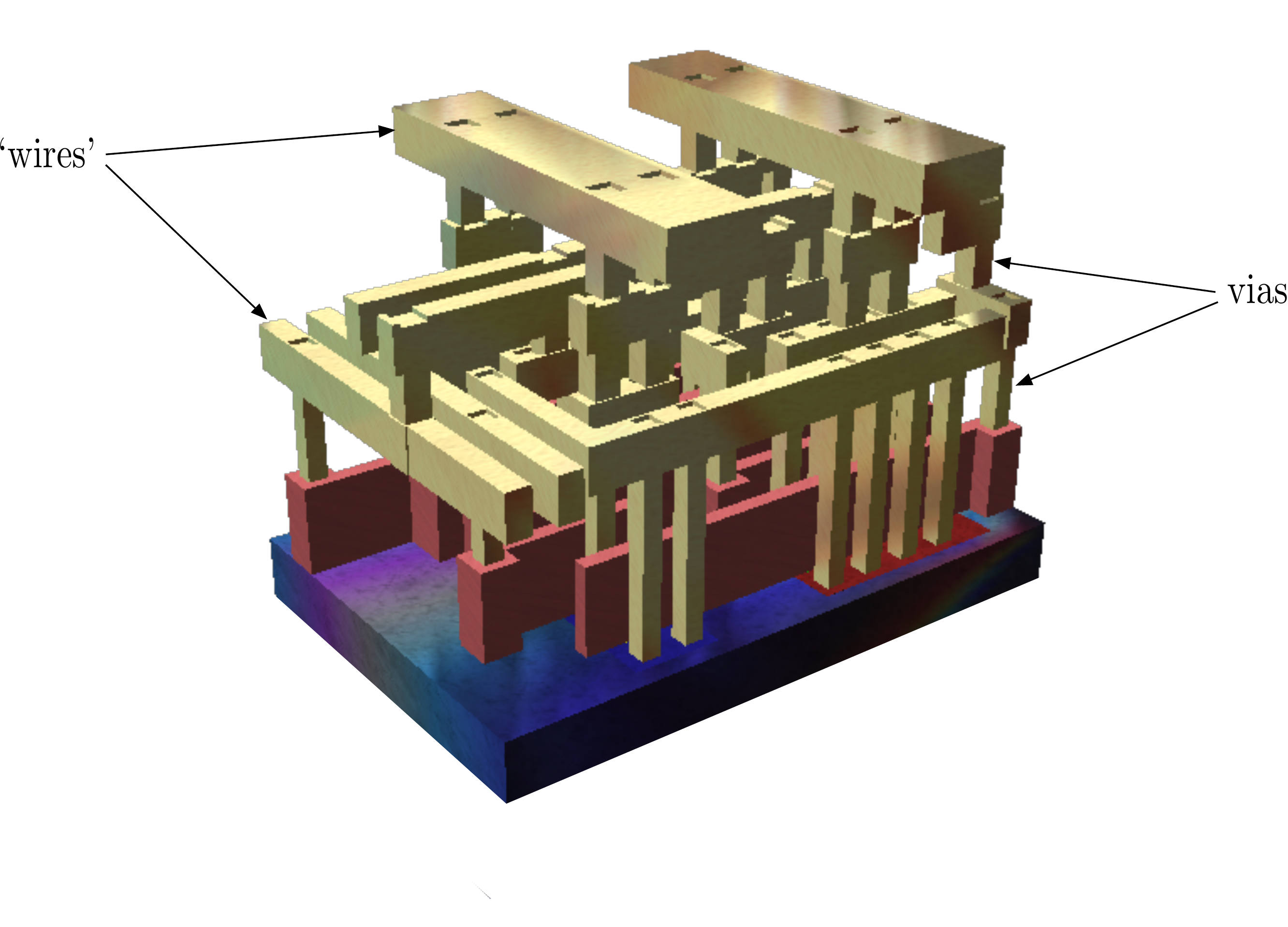} &  \includegraphics[scale = 0.27]{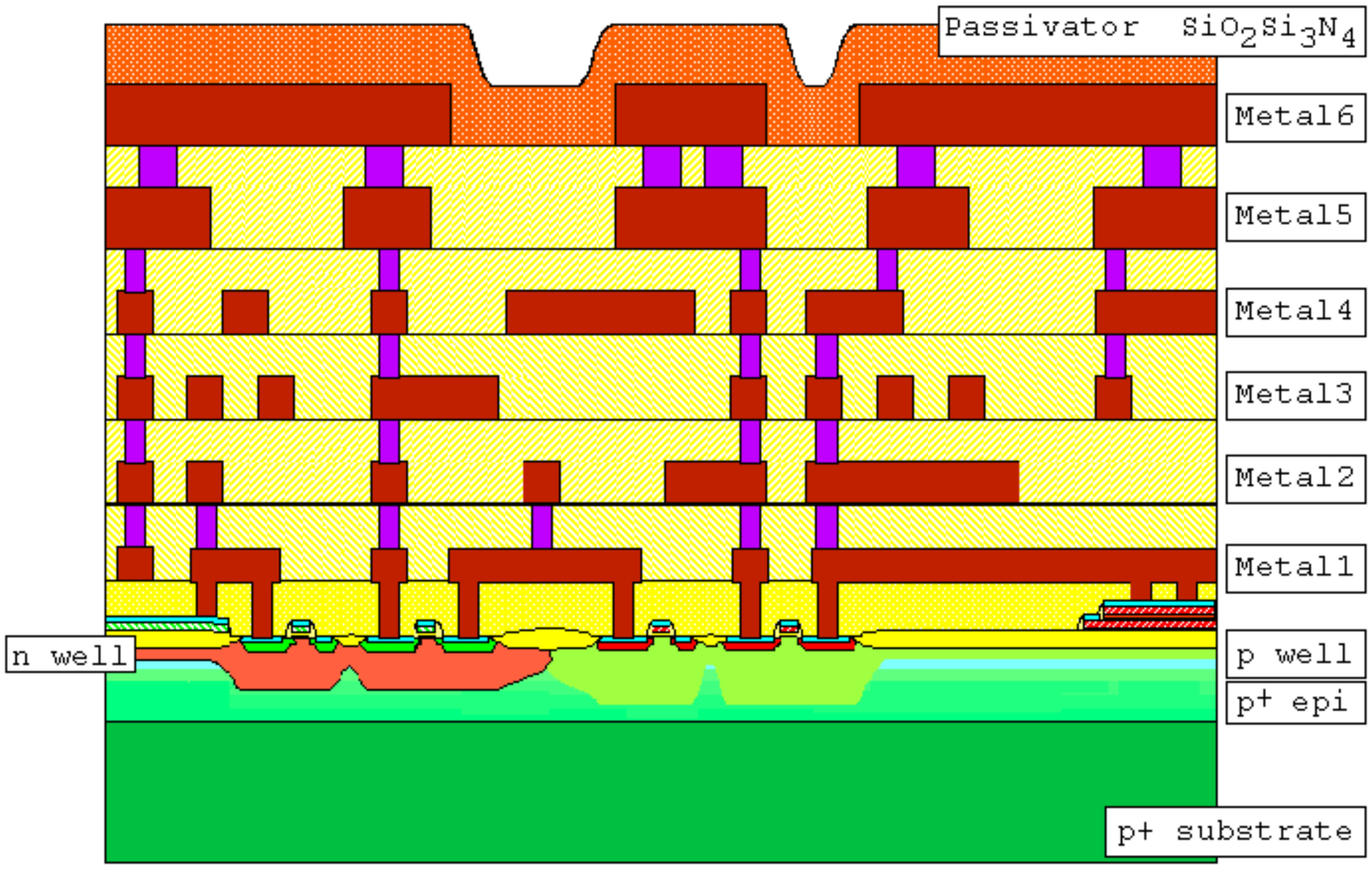}\\
	(a) & (b) \\
	\end{tabular}
	\caption{(a) a 3D view of an integrated circuit (source: \url{https://commons.wikimedia.org/wiki/File:Silicon_chip_3d.png}); (b) a cross-section of an integrated circuit: the first layers represent the substrate and the transistors, the dark red components represent metal `wires', and the purple components represent vias.}
	\label{integrated_circuit}
	\end{center}
\end{figure}

During the fabrication of integrated circuits, a large number of transistors are etched over a {\em silicon wafer} (or silicon substrate). Then, a dense network of metal conductors is deposited on multiple layers within the dielectric material (non-conductive medium) on top of the transistors. The network provides the electrical current paths among the different components (see Fig. \ref{integrated_circuit} for illustrations of integrated circuits). As illustrated in Fig. \ref{integrated_circuit}(a), the layers are typically of two kinds: either they contain (non-crossing) segments or snake-like shapes that somehow connect components horizontally - the corresponding metal shapes are called {\em wires}, and we refer to such layers as {\em metal layers} - , or they contain vertical square cylinders that allow connecting successive metal layers -  the corresponding metal shapes are called {\em vias}\footnote{Observe that we might superimpose several square cylinders in consecutive (metal and via) layers so that the corresponding metal component connects non-successive metal layers. The corresponding component is also referred to as a via in the industry.}, and we refer to such layers as {\em via layers}.  {\em DSA-aware multiple patterning} combines lithography and Directed Self-Assembly technologies. We now detail the two technologies and the corresponding process.

\subsubsection*{Lithography}

\begin{figure}
\begin{center}
\begin{tabular}{ccc}
\includegraphics[scale=0.25]{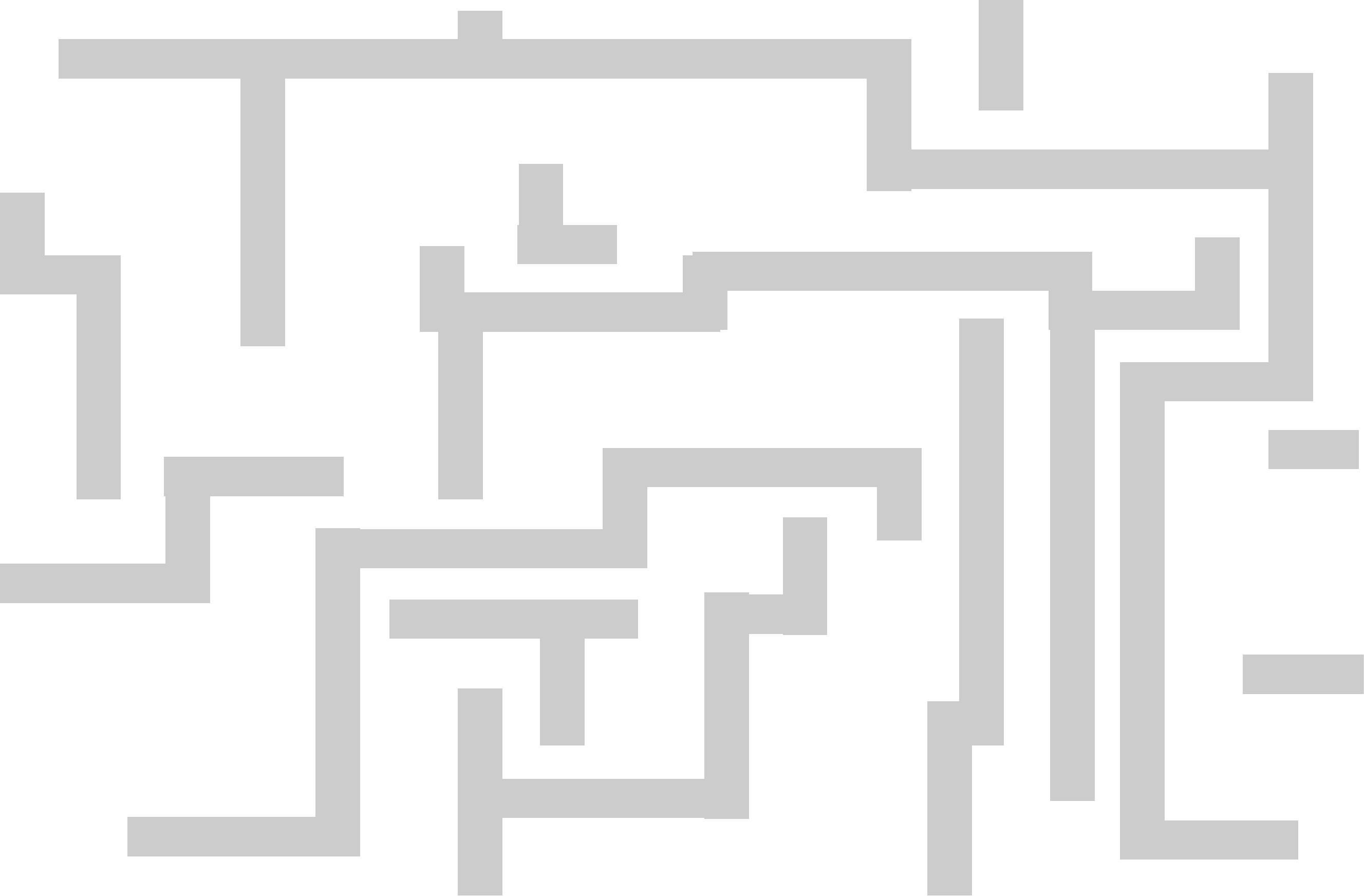} & \hspace{2cm} & \includegraphics[scale=0.25]{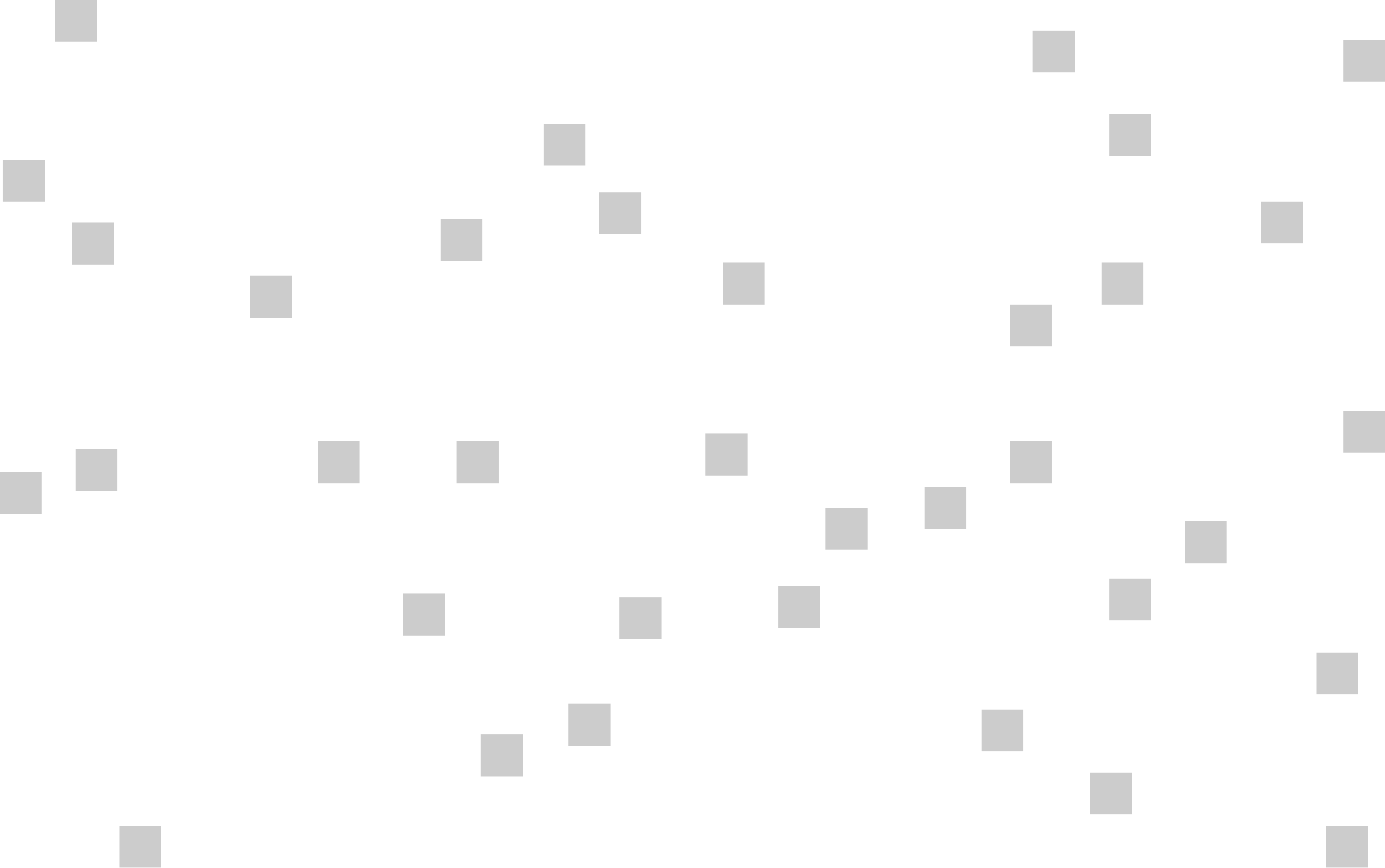} \\
(a) & & (b)\\
\end{tabular}
\end{center}
\caption{Two examples of a layout: (a) a metal layer; (b) a via layer}
\label{layout}
\end{figure}

\begin{figure}
	\begin{center}
	\includegraphics[scale = 0.3]{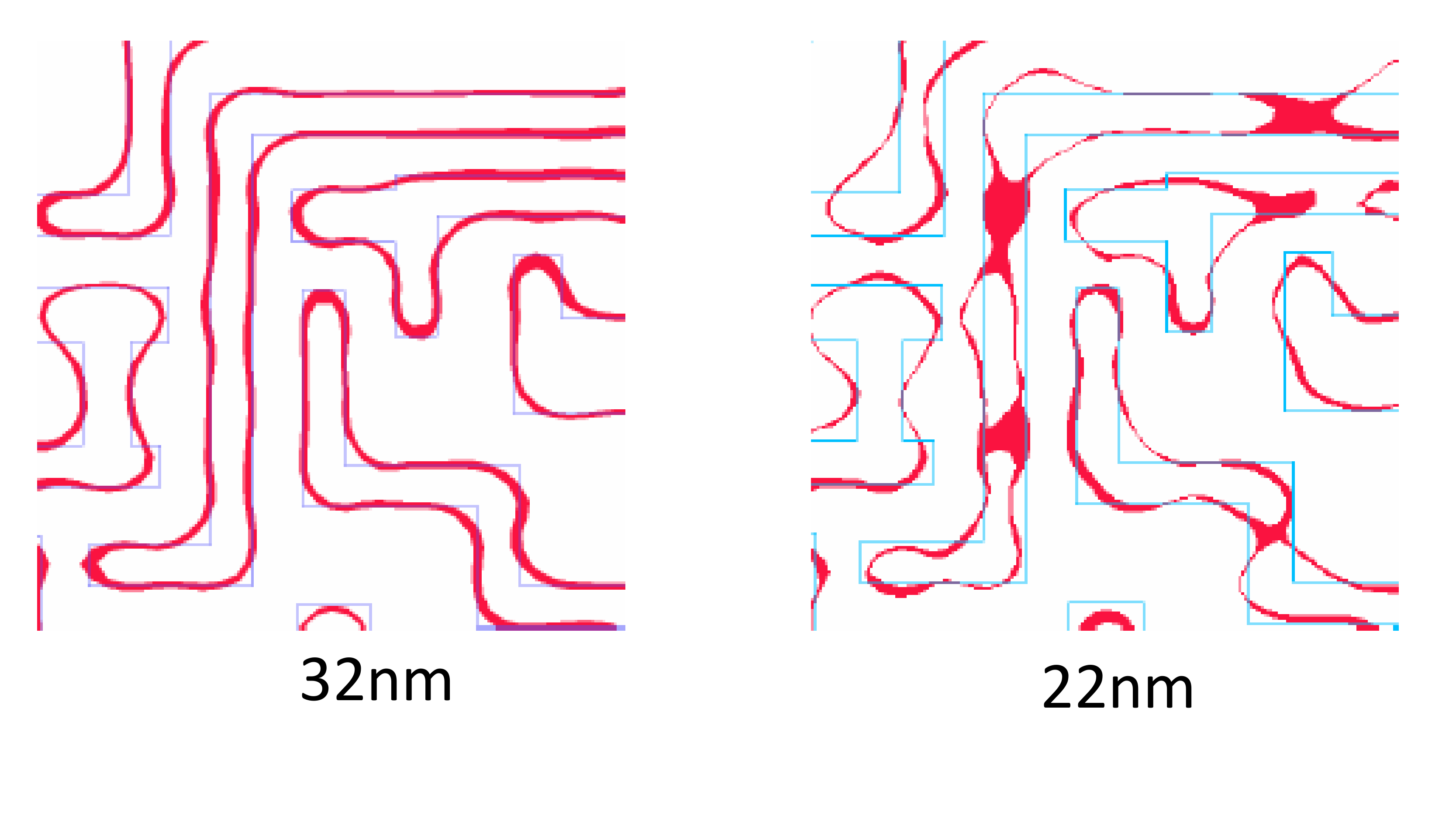}
	\caption{Comparing single patterning between 32nm and 22nm processes: as the resolution decreases, optimal distortion might induce defects. }
	\label{32vs22}
	\end{center}
\end{figure}

Lithography is typically used to `transfer' geometrical {\em features} (vias, segments, or other objects) of the same layer from a mask to the wafer. This is achieved by exposing a light-sensitive chemical photoresist that is deposited on the wafer to a light source through the mask. This creates a `mold' that can later be `filled' with a conductive material through various chemical operations, a process called {\em etching}. The arrangement of {features} to be transferred is usually referred to as a {\em layout}. Fig.~\ref{layout} shows two different examples of a layout (viewed from the top). Note that in our illustrations, we usually draw `idealized' shapes for the features. In particular, we adopt the Electronic Design Automation (EDA) convention of representing vias as squares (we also assume that the squares are of equal size within a given via layer, which is also common practice). If the same shapes were used on the mask, the final shapes on the silicon wafer would differ due to optical distortions, which may depend on the technology used, but the transferred shapes are typically more rounded (see Fig. \ref{32vs22}) so that a square on the mask generates a squircle (rounded square), or for a more advanced technology node, a circle on the wafer. As long as the network structure is preserved, the precise shape of the features on the wafer does not matter much. Optimal proximity correction (OPC) might be used to adjust the shapes on the mask upfront to ensure that the final arrangement is as close as possible to the targeted one. In addition, functional tests are performed on the final circuit to verify that manufacturing was successful. The reader can refer to \cite{Dehia} for a more detailed overview of the whole lithography process.

Optical distortions could induce network defects, such as the disruption of a wire or the `fusion' of several wires (see Fig. \ref{32vs22}). This occurs when the features are too close to each other. The minimum distance permitted between any two features to prevent defects is usually referred to as {\em lithography distance (or resolution)}, which we denote with $Litho_{dist}$ (note that the distance we consider between two features $f_1, f_2$ is the Euclidian distance i.e., $\min_{x\in f_1,y\in f_2} ||x-y||_2$, that is, the distance border to border). 
For instance, 193 immersion technology has a resolution limit of 45nm \cite{193immersion}, while the next generation of lithography, based on Extreme Ultra-Violet (EUV) light, allows lowering the resolution to 27nm \cite{EUVresolution}. As EUV is currently not used at a large production scale, and lithography technologies have tended to reach their limit, the industry is seeking other solutions to further lower the resolution (currently targeting sub-7 nm resolution). As pointed out in Section \ref{introduction}, multiple patterning is one such solution.

Multiple patterning is conceptually simple: the idea is to decompose the original layout into feasible sub-layouts that will be etched with different masks, one after the other, to produce the original arrangement. While multiple patterning may potentially decrease the minimum possible distance between the features within a layer, it substantially increases production costs and time. Indeed, masks are  expensive, and given the fact that modern integrated circuits might contain fifteen to twenty layers (and typically involve around fifty rounds of lithography), the cost of all masks needed to manufacture an integrated circuit could reach millions of dollars. Furthermore, one of the main drawbacks of multiple patterning is alignment. For instance, in the case of double patterning, first, the features of the first sub-layout are etched on the silicon wafer. Then, the features of the second sub-layout have to be aligned with the first set of printed features and etched on the silicon wafer. When the number of patterning steps increases, the perfect alignment of features from different masks becomes challenging. Together with the reduction in the throughput discussed in Section \ref{introduction}, these are the main reasons why the number of patterning steps used in the industry is usually kept small.\\

The current standard in manufacturing is in fact to use double patterning (DP) in most cases, and then triple patterning (TP) or quadruple patterning (QP) when DP is not feasible. Quadruple patterning allows managing most (current) practical situations, but again, due to increased production costs and time, the industry is seeking solutions to minimize the number of patterning steps in the production of each layer. The corresponding problem readily translates into a graph coloring problem. Indeed, consider the graph whose node set are the features and where two nodes are adjacent if the distance between the corresponding features is below the lithography distance. This graph is usually called the {\em conflict graph}. Minimizing the number of patterning steps is equivalent to finding the chromatic number of this graph, that is, the minimum number of colors needed to color the vertices of the graph such that no pair of vertices within the same color are adjacent (a coloring with this latter property is usually called {\em proper}).\\

Proper (vertex) coloring is a notoriously NP-hard problem \cite{Karp} and testing whether a graph can be colored with a fixed number $k\geq 3$ of colors is NP-complete \cite{Stockmeyer}. The problem can be solved in polynomial time for some very specific classes of graphs, such as perfect graphs \cite{GLS} or graphs with bounded tree-width \cite{Bodlaender} for instance, but usually remains hard even when additional assumptions are made on the graph structure (for a recent survey of complexity results and algorithms for graph coloring, see \cite{Golovach}). As conflict graphs arising from manufacturing an integrated circuit have some structure, it is natural to wonder whether this allows for polynomial time algorithms. For instance, as conflicts arise from proximity, when restricting to the manufacturing of vias, and if assuming that we are fine\footnote{This is usually the case.} with producing (equal size) cylinders - and not squares - the corresponding graphs are unit disk graphs. Unfortunately, the problem remains hard in this class of graphs (even for planar unit disk graphs and also simply checking 3-colorability) \cite{Peeters}. Some authors have studied other types of structures that might be relevant \cite{grid}. Computational complexity and exact and heuristic approaches for proper vertex coloring have been surveyed in \cite{Pardalos} and \cite{Malaguti2}. For additional references on exact approaches, see also \cite{Campelo,Hansen,Mendez,Mehrotra,Gualandi,Malaguti}. Furthermore, from the application side, several exact and heuristic approaches have been developed, see \cite{SurveyMPL} for a survey. \\

There is strong industrial interest in new processes that can be used on top of multiple patterning to further reduce the number of patterning steps. {\em Directed Self Assembly} (DSA) has been identified as a promising solution in the manufacturing of vias, as other alternative techniques such as stitching are not applicable in this context \cite{Ma}. 


\subsubsection*{Directed Self Assembly (DSA)}

DSA is a chemical approach based on {\em block copolymers (BCP)} -- a combination of two different structures (i.e., attraction between different molecules) -- that works as follows: a region, called a {\em guiding pattern}, is filled with BCP in a `random' state (i.e., an unorganized mixture of different blocks of molecules). After a certain chemical reaction is triggered (called a {\em microphase separate anneal}), the BCP assembles into a periodic arrangement of homopolymer structures:  the periodic structures can be cylinders, lamella, or other geometric structures. These structures depend on the nature of the block copolymer and the {\em volume fraction} (the ratio of volume occupied by the two homopolymers). In this work, we are interested in the periodic cylinder structures as they might be used as vias. Indeed, one can combine microphase separate with additional chemical steps to retain the negative of cylinders. \\

\begin{figure}
	\begin{center}
	\includegraphics[scale = 0.5]{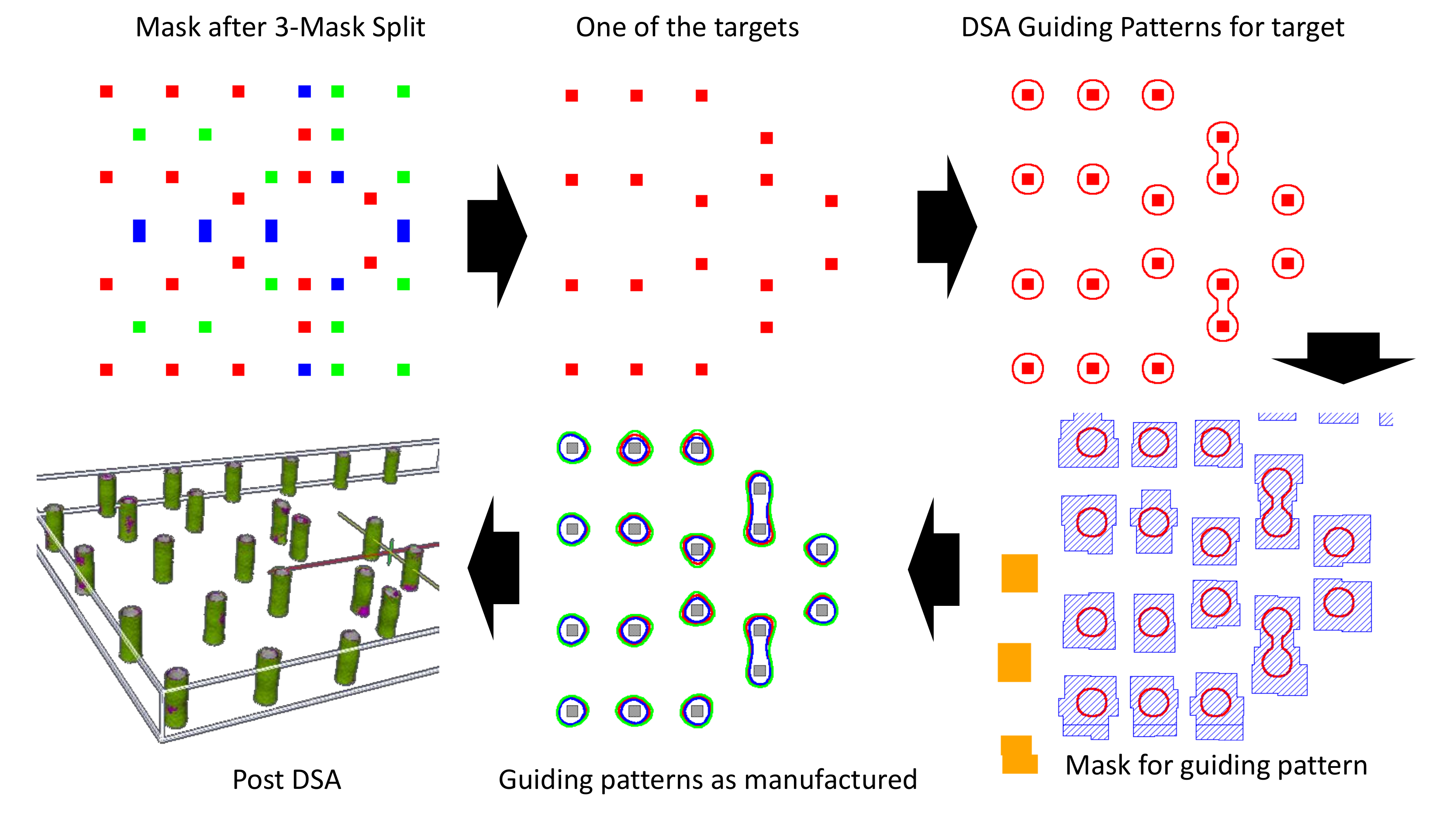}
	\caption{Example of DSA-aware Triple Patterning. The layout is decomposed into three sub-layouts (red, blue, green). We detail the manufacturing of the red sub-layout using DSA: (i) there are two pairs of vias that are in conflict in the red sub-layout (we do not provide the resolution here but it can be selected as precisely the smaller distance between any two vias in the red sub-layout); (ii) the corresponding pairs are grouped into peanut-like guiding patterns that will be used to direct the assembly of the block copolymer; (iii) the associated mask is then created, taking into account optical distortion; (iv) the guiding patterns for the red sub-layout are then manufactured through one lithography step; (v) finally, the vias are etched through DSA.   }
	\label{DSAflow}
	\end{center}
\end{figure}

Such a process can readily be combined with lithography to reduce the number of patterning steps in the manufacturing of vias. The whole idea in mixing DSA and lithography is to group some vias into guiding patterns that could otherwise not be assigned to the same mask. Lithography is then used to `mold' the guiding patterns, and DSA is used to etch the vias that lie within these patterns (see Fig. \ref{DSAflow} for an example). We distinguish two kinds of masks in this process: we call {\em DSA mask} a mask that involves a {\em non-trivial} guiding pattern (at least two vias are grouped in this mask), and {\em Litho mask} a mask that does not involve guiding patterns. Manufacturing constraints impose that a DSA mask can only use one block copolymer to etch vias within that mask: indeed all vias in the mask are etched through DSA after all guiding patterns have been printed through lithography (even single vias of this mask will be printed with DSA and will thus appear as cylinders on the wafer). In this study, we additionally assume that we only use one block copolymer for all masks. The production costs of this new process are again dominated by the cost of the masks, and production throughput is again limited by the number of patterning steps. Hence, it is still essential to minimize the number of lithography/patterning steps. \\

Heuristics and exact approaches for multiple patterning with DSA (and variants) have been investigated in \cite{Badr,Fang,Kuang, Ou,Ou2,Yang}. Note that in all studies, the number of patterning steps is fixed and the goal is to group vias into feasible guiding patterns so as to minimize the number of conflicts remaining (allowing sometimes for the insertion of redundant vias). In contrast, this work focuses on the ``pure'' coloring problem, that is, explicitly finding the minimum number of patterning steps needed for manufacturing with DSA (with no conflict allowed). This is motivated by two different goals: the first to formally demonstrate the potential benefits that DSA-aware multiple patterning could bring (over pure multiple patterning), and the second to allow assessing the quality of the heuristics developed in-house by Mentor Graphics.

In principle, it would be possible to use the exact methods developed in some prior studies in parallel: run the algorithm for a given number of patterning steps and then verify for which number of patterning steps zero coloring conflicts emerge. However, most of these methods either employ heuristics to accelerate finding a coloring solution at a large scale (and hence no longer guaranteeing optimality) \cite{Ou,Yang}, propose formulations that do not work for any number of patterning steps \cite{Badr, Kuang}, or exploit additional structures and/or placement options \cite{Fang,Kuang,Ou2}. In our case, given that the objective is to formally find the minimum number of patterning steps required for (large scale) layouts, we do not build on the methods developed in prior research.

Beyond the relevance from a practical point of view, we believe that our new models deserves additional attention from the combinatorial optimization community, as they are natural extensions of proper graph coloring and may find other applications beside integrated circuit manufacturing.

\section{Relation to graph coloring and IP formulations}

There are several natural ways of exploiting DSA within Multiple Patterning. We now detail a few variants that are of particular interest to the industry, their relation to graph coloring problems, and some `natural' integer programming formulations. Note that we essentially extend the standard assignment-based integer programming formulations for vertex coloring (of course, the generalization brings other complications). We will now explain the rationale behind this choice. While it is known that the corresponding model contains color symmetries, and that the corresponding linear relaxation is weak \cite{Mutzel, Malaguti2}, it has the advantage of being easily implementable in modern solvers such as CPLEX or GUROBI. In our setting, because the upper bound on the number of colors is small, the color symmetries are limited, and therefore not very problematic (actually, we undertook some preliminary tests with column generation approaches in BaPCod \cite{BapCod}, and the `assignment' formulations implemented with Cplex 12.6.3 always performed better; indeed, experts of decomposition techniques \cite{Clautiaux} confirm that this is not surprising as the sub-problem is `as hard' as the original one when the chromatic number is small). Furthermore, many specific cuts, such as clique inequalities for instance, are available in these solvers so we can also easily strengthen the formulation by simply activating well-known strong cuts for the problem (cliques, Chvatal-Gomory cuts, etc...).

\subsection{Pairing vias}\label{pairing}

The first obvious idea to exploit DSA together with lithography is to attempt to group vias by pairs. As known \cite{Badr}, two vias can be grouped if they stand within a distance in a range of $[L_0,U_0]$ (center to center), which depends on the BCP, and if they satisfy additional constraints based the lithography technology (for instance, in 193 immersion, the contours of the guiding patterns have to be parallel to the x and y axis). 

In this case, minimizing the number of patterning steps in DSA-aware Multiple Patterning is a simple variant of graph coloring. Let $G=(V,E)$ be the conflict graph associated with the chosen lithography technology. Let $F\subseteq E$ be the set of edges of $E$ whose extremities are within a distance between $L_0$ and $U_0$, and satisfy the additional lithography constraints associated with the technology (we sometimes call such edges {\em DSA edges}). The problem is coloring the vertices of $G$ with a minimum number of colors so that each color induces a disjoint union of nodes of $G$ and edges of $F$, or, alternatively, each color induces a graph where all nodes have at most $1$ degree and all edges are in $F$\footnote{This assumes that guiding patterns are in conflict with other guiding patterns if and only if some of the corresponding vias are in conflict, which is a reasonable assumption according to the industry given the typical values of $L_0, U_0$ and $Litho_{dist}$.}. When $F=E$ the problem is known as 1-improper coloring and is NP-hard \cite{Havet}. In fact, according to the same authors, it is already hard to check whether a graph admits a 1-improper 2-coloring. To the best of our knowledge, the problem has not yet received much attention from the combinatorial optimization community. However, as known, constant factor approximation algorithms exist (see again \cite{Havet}). The problem can easily be formulated as an integer program, building on the standard graph coloring formulation, as follows ($L$ is an upper bound on the number of colors - in practice, because most designs can be solved with quadruple or quintuple patterning, $L$ can be set to $4$ or $5$, as a proper coloring is obviously 1-improper)\footnote{We denote with [L] the set $\{1,...,L\}$ and with $N_F(u)$ the neighbors of $u$ in the subgraph $G_F:=(V,F)$.}:

\begin{align}
\min & \sum_{i=1}^L \lambda^i \label{induced_02_obj}\\
	& \sum_{i=1}^L z^i_v 				&& = \quad 1, 				&& \forall i \in [L], \forall v\in V  \label{induced_02_cst1}\\
	&  z^i_u + z^i_v -1 		&& \leq  \quad x^i_{(u,v)}, && \forall i \in [L], \forall (u,v)\in F \label{induced_02_cst2}\\
	& \sum_{v\in N_F(u)} x^i_{(u,v)} && \leq \quad 1, 		&& \forall i \in [L], \forall u\in V  \label{induced_02_cst3}\\
	& z^i_u + z^i_v 			&& \leq \quad 1, 			&& \forall i \in [L], \forall (u,v)\in E\setminus F  \label{induced_02_cst4}\\
	& z^i_u, x^i_{(u,v)} 		&& \leq \quad \lambda^i 		&& \forall i \in [L], \forall u \in V, \forall(u,v) \in F   \label{induced_02_cst5} \\
	& z^i_u, x^i_{(u,v)}, \lambda^i 		&& \in \quad \{0,1\},	&& \forall i \in [L], \forall u \in V, \forall(u,v) \in F
\end{align}

  Variable $\lambda^i$ indicates whether color $i$ is used, $z^i_u$ indicates whether vertex $u$ is assigned color $i$, and $x^i_{(u,v)}$ indicates whether edge $(u,v)$ belongs to color $i$ (that is, with both extremities in color $i$). Constraint (\ref{induced_02_cst1}) ensures that each vertex is colored. Constraint (\ref{induced_02_cst2}) ensures that if an edge of $F$ is not selected within color $i$, then the extremities cannot both receive color $i$. Constraint (\ref{induced_02_cst3}) ensures that no vertex of color $i$ is adjacent to more than one other vertex within that color (through an edge of $F$).  Constraint (\ref{induced_02_cst4}) ensures that there is no conflict within a color. Finally, (\ref{induced_02_cst5}) ensures that vertices and edges are assigned to a color only if the color is selected. The number of constraints and the number of variables in this formulation are in the order of $O(L.n^2)$, where $n$ is the number of nodes of the graph.


\subsection{Small groups}\label{small}

In principle, it is possible to group more than two vias within guiding patterns. Indeed, design rules for guiding patterns have been investigated with explicit constraints on feasible groups in \cite{Xiao} and \cite{Chang}. However, for the time being, there are only few specific shapes of guiding patterns that are validated. Furthermore, as the guiding patterns will have to be etched using lithography, the lithography technology will also have an impact on the feasible groups (as in the case of pairing, see above). The feasibility of guiding patterns can be verified through a procedure called {\em DSA flow}. If we assume that we are given a complete list ${\mathcal V}\subseteq 2^V$ of all feasible groups (including singletons) and a complete list $\mathcal E$ of all pairs of $\mathcal V$ in conflict (in particular, two groups containing the same via will be in conflict), we can model the problem as another variant of graph coloring. Let $\mathcal G$ be the graph with vertex set $\mathcal V$ and edge set $\mathcal E$. We want to find a subset $\mathcal U$ of groups of $\mathcal V$ satisfying $\bigcup_{g\in \mathcal U} g= V$ with $\chi(G[U])$ (the chromatic number of $G[U]$) minimum.  

Of course, we might consider variants of this problem where $\mathcal V$ is a subfamily of feasible groups that have been validated, such as pairs of vias, for instance. In practice, a limited number of vias can be grouped due to manufacturing constraints. The maximum number of vias per group might evolve in the future, but with current technology is typically limited to two or three. This gives rise to the following mixed integer program ($L$ is again an upper bound on the number of colors):

\begin{align}
\min & \sum_{i=1}^L \lambda^i  \\
	&\sum_{i=1}^L \sum_{g \in \mathcal V: v\in g} x^i_g 		&& = \quad 1,	&& \forall v\in V \label{naive_cst1}\\
	&x^i_{f} + x^i_{g} 								&&\leq \quad 1, && \forall i \in [L], \forall  (f, g) \in \mathcal E \label{naive_cst2}\\
	&x^i_g 												&&\leq \quad \lambda^i, 	&& \forall i \in [L], \forall g\in {\mathcal V} \label{naive_cst3}\\
	&x^i_g, \lambda^i 									&& \in  \quad \{0,1\}, 		&& \forall i \in [L], \forall g\in {\mathcal V}\\
\end{align}

Variable $\lambda^i \in \{0,1\}$ indicates whether color $i$ is chosen, and $x^i_g \in \{0,1\}$ indicates whether a group $g \in \mathcal{V}$ is colored with color $i$.
Constraint (\ref{naive_cst1}) imposes that each node $v \in V$ is assigned to exactly one group and one color. Constraint (\ref{naive_cst2}) imposes that two groups $f$ and $g$ in conflict have to receive different colors.  

In this model, the number of variables is in the order of $O(L.n^k)$ and the number of constraints is in the order of $O(L.n^{2k})$, where $n$ is the number of nodes of the (original) graph and $k$ is the maximum number of vertices allowed in a group. In our practical applications, we dealt with groups of size two or three. In this case, such a na\"ive enumerative approach appears to perform pretty well, as we will see in Section \ref{experiments}.

\subsection{Larger groups}\label{large}

When the maximum size $k$ of the groups increases (even if still bounded - consider $k=6$ for instance), the previous model would quickly become too large to be handled by a modern solver for practical size instances, as the number of variables and the number of constraints grow exponentially in $k$. It is tempting in this case to try to develop models that avoid the enumeration of the feasible groups and instead build the optimal groups together with the coloring. To develop integer programming models in this case, we must understand and exploit the structure of the groups. 

The main certified feasible groups put forth in \cite{Chang,Bekaert} concern {\em `paths' of vias}. We focus on this special case, as this is what industrial companies are currently mainly interested in. We also assume that we have a bound $k$ on the number of vias in the paths (see the discussion in Section \ref{small}).

As discussed in Section \ref{pairing}, vias at distance in $[L_0,U_0]$ can be paired. In fact, under some additional conditions on the path obtained, they can be `chained'. More formally, let $G=(V,E)$ be the conflict graph associated with the chosen lithography technology. Let $F\subseteq E$ be the set of edges of $E$ whose extremities are at distance (center to center) in $[L_0,U_0]$. Manufacturing constraints allow associating feasible groups with induced paths of length $k-1$ (the length of a path counts the number of edges) in the subgraph $G_F=(V,F)$ as long as it complies with the constraints associated with the lithography technology used. For instance, in 193 immersion, the paths have to be parallel to the x or y axis, and in EUV, the angle (degree) between any three consecutive vias in the paths should be in the range of $[135,225]$. 

If we ignore the lithography-specific restrictions (we can easily add the corresponding restrictions later in the integer programming model, see Section \ref{experiments}), the problem is yet another variant of graph coloring that can be described as follows. Given a graph $G=(V,E)$, $F\subseteq E$, and an integer $k\geq 1$, color the nodes of $G$ so that each color induces a disjoint union of paths of length at most $k-1$, using only edges of $F$. 


When $k=1$, the problem is a standard graph coloring problem. When $k=2$, the problem only allows pairs and is thus closely related to the 1-improper coloring problem. For larger values of $k$ and when $F=E$, the problem was introduced by \cite{Akiyama} as the $(k-1)$-path coloring problem\footnote{Note that some authors use the same terminology for another variant of graph coloring, see for instance \cite{Frick,Johns,Mynhardt}.}. The question of the $(k-1)$-path $L$-colorability of a graph was coined as the $(k,L)$-path coloring problem in \cite{Jinjiang2}. Jinjiang proved that the $(2,2)$-path coloring problem and the $(3,3)$-path coloring problem are NP-complete \cite{Jinjiang2,Jinjiang}. Thus, the $1-$path $2-$coloring and the $2-$path $3-$coloring problems are already NP-complete. We again develop a natural integer programming formulation for the problem when $k\geq 2$ (for $k=1$, we can use the standard coloring formulation).

\begin{align}
min & \sum_{i=1}^L \lambda^i \\
	& \sum_{u\in N_F(v)}x^{i,\kappa}_{(v,u)} 										&& \leq \quad y^i_{v}, && \mbox{ if } \kappa=0, \forall i \in [L], \forall v \in V \label{induced_04_cst1}\\
	& \sum_{u\in N_F(v)}x^{i,\kappa}_{(v,u)}-\sum_{u\in N_F(v)}x^{i,\kappa-1}_{(u,v)} && \leq \quad 0, && \forall \kappa: k-2 \geq \kappa \geq 1, \forall i \in [L], \forall v \in V \label{induced_04_cst2}\\
	&  \sum_{\kappa=0}^{k-2} (x^{i,\kappa}_{(u,v)} + x^{i,\kappa}_{(v,u)})			&& =  \quad x^i_{(u,v)},	&&  \forall i \in [L], \forall (u,v)\in F \label{induced_04_cst3} \\ 
& y^i_{v}+\sum_{\kappa=0,...,k-2} \sum_{ u \in N_F(v)}x^{i,\kappa}_{(u,v)} 		&& = \quad z^i_v,			&&  \forall i \in [L], \forall v\in V \label{induced_04_cst4} \\
& \sum_{i=1}^L z^i_v && = \quad 1, && \forall v\in V \label{induced_04_cst5} \\
&  z^i_u + z^i_v -1 && \leq  \quad x^i_{(u,v)}, && \forall i \in [L], \forall (u,v)\in F  \label{induced_04_cst6} \\
& z^i_u + z^i_v && \leq \quad 1, && \forall i \in [L], \forall (u,v)\in E\setminus F \label{induced_04_cst7}\\
& z^i_u, x^i_{(u,v)} && \leq \quad \lambda^i 	&&\forall i \in [L], \forall u \in V, \forall(u,v) \in F \label{induced_04_cst8}\\
& z^i_u, x^i_{(u,v)}, y^i_{v}, x^{i,\kappa}_{(u,v)}, \lambda^i && \in \quad \{0,1\}, && \forall i \in [L], \forall \kappa=0,...,k-2,  \forall u,v \in V, \forall(u,v) \in F
\end{align}

Variable $x^i_{(u,v)}\in \{0,1\}$ indicates whether the edge $(u,v)\in F$ is assigned to color $i$.  $x^{i,\kappa}_{(u,v)} $ and $x^{i,\kappa}_{(v,u)}\in \{0,1\}$ indicate whether the edge $e=(u,v)\in E$ is used as the $(\kappa+1)$-th edge in the direction from $u$ to $v$  or $v$ to $u$ in one of the disjoint paths of color $i$ (explicitly giving an orientation to the path). $y^i_{v}\in \{0,1\}$ indicates whether there is a path that `starts' from $v$ in color $i$ (it might be a path of length 0). $z^i_v \in \{0,1\}$ indicates whether a node $v \in V$ has color $i$ (hence $v$ is in a path of color $i$, possibly of length 0). Finally, $\lambda^i \in \{0,1\}$ indicates whether color $i$ is chosen. 

Constraints (\ref{induced_04_cst2}) and (\ref{induced_04_cst1}) are `flow conservation constraints' that impose that an edge leaving from $v$ (in color $i$) can be the $(\kappa+1)$-th edge of a path only if there is an edge entering $v$ that is the $\kappa$-th, and that there cannot be a path starting with an edge from $v$ unless $v$ is the first node of the path. Constraint (\ref{induced_04_cst3}) imposes that $(u,v)$ is taken in color $i$ if and only if it is used in one direction or the other in a path. Constraint (\ref{induced_04_cst4}) ensures that a vertex $v$ in color $i$ is either the `starting' extremity of a path of color $i$ or $u\in N_F(v)$ exists such that the edge $(u,v)$ is taken in a path of color $i$ in the direction from $u$ to $v$ (and vice versa). Constraint (\ref{induced_04_cst5}) guarantees that each vertex receives a color. Constraint (\ref{induced_04_cst6}) ensures that if an edge of $F$ is not selected within color $i$, then the extremities cannot both receive color $i$. Constraint (\ref{induced_04_cst7}) ensures that there is no conflict within a color. The number of constraints and the number of variables in this formulation are in the order of $O(L.k.n^2)$, and the number of constraints in the order of $O(L.n^2)$, where $n$ is the number of nodes of the graph and $k$ is the number of nodes in the path. One of the main advantages of this formulation is that it grows linearly in $k$, and could thus, in principle, be implemented in modern solvers for larger values of $k$ than the previous model. However, it is less flexible as it is limited to paths, and as we will see later, is much weaker. 




\subsection{Beyond induced paths}\label{general}

Requiring that paths be induced is somewhat conservative: for instance, three vias that are aligned, whose middle node is at a distance $L_0$ from each extremity, and where the two extremities are in conflict, might qualify for possible grouping (since the corresponding guiding pattern would in principle allow for the proper assembly of the three vias according to \cite{Chang}). Hence, while induced paths are guaranteed to correspond to feasible groups, other paths might be allowed. However, in practice, it seems that the distances are often such that the situation described above for three vias does not emerge ($Litho_{dist}$ is `not too big' compared to $L_0$), and in the case of 193 immersion in particular, preventing this `three vias case' is enough to ensure that all feasible paths (i.e., parallel to the x or y axis) are actually induced. When testing our model on true instances (see the next section), permitting non-induced paths did not allow better solutions. However, we believe that the relation between $L_0$ and $Litho_{dist}$ might evolve in the future and that studying more general models makes sense. \\

\begin{figure}
    \centering
    \includegraphics[scale=0.5]{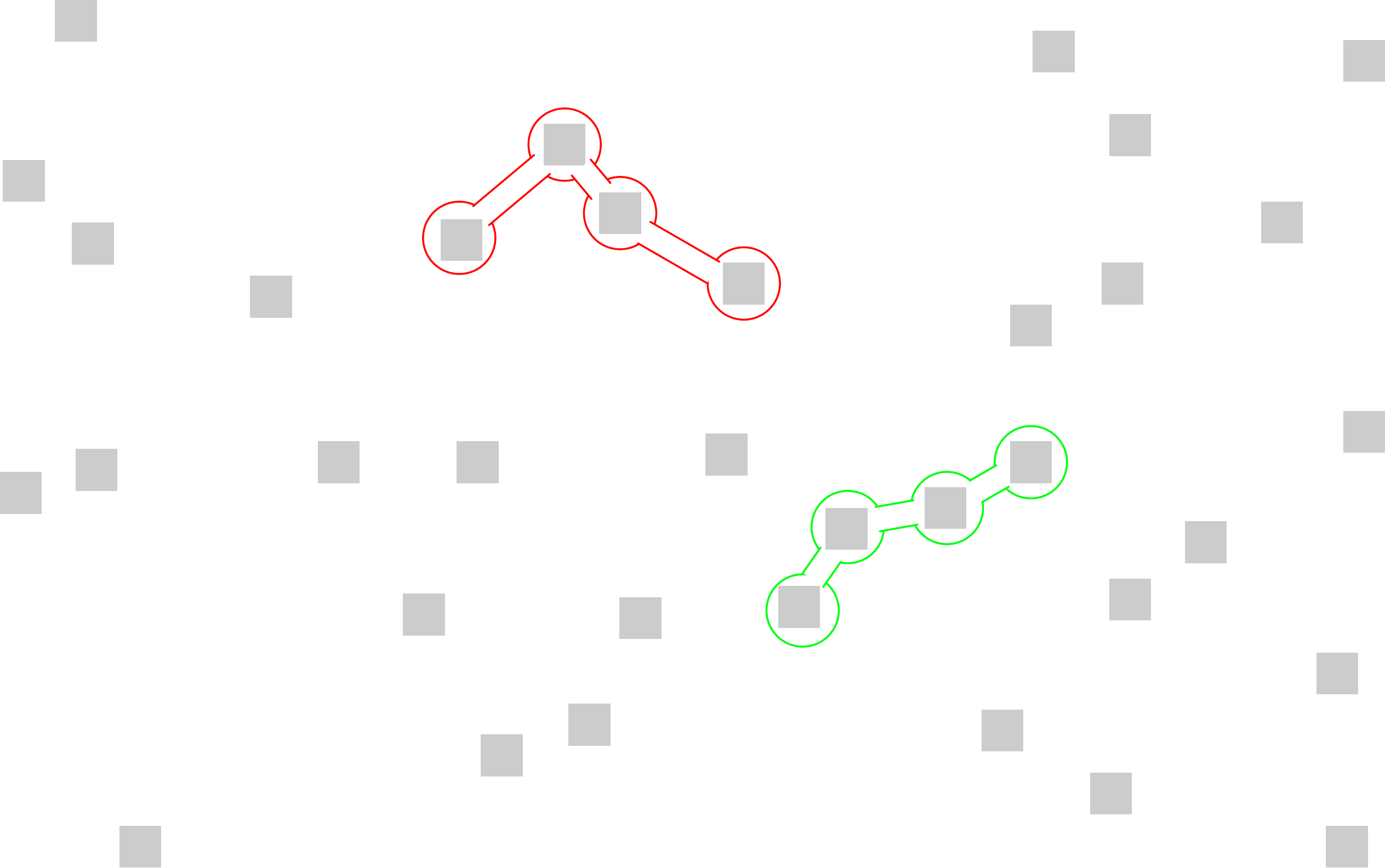}
    \caption{Assume that $Litho_{dist}$ is such that all vias within the green and the red guiding patterns are in conflict. The green guiding pattern would be fine, as it is `close' enough to a straight line and would thus not be greatly affected by optical distortion, while the red guiding pattern would certainly induce defects. }
    \label{non_induced}
\end{figure}

A natural relaxed assumption is to require that a set $U\subseteq V$ with at most $k$ vias can be grouped if there is a Hamiltonian path in the subgraph $G(U,F)$. The existence of the Hamiltonian path ensures that we can create, in principle, a guiding pattern that closely follows the path that might assemble properly. Of course again, lithography might additionally impose some constraints on the guiding pattern. For instance, one might want to impose that, within a guiding pattern, there are no two vias $v_1$ and $v_2$ that are in conflict, and such that the segment linking these two is not `close'  to the path that links $v_1$ and $v_2$ in the Hamiltonian path (otherwise, the position of the vias would certainly differ from what is expected since the guiding pattern might itself differ substantially from the targeted one due to optical distortion): a natural measure of proximity might be to impose that each vertex of the path linking the two vias should be within a certain maximum Euclidean distance from the segment (see Fig. \ref{non_induced} for an example). For 193 immersion, this latter restriction is granted once we impose that the paths are parallel to the axis. For other technologies, such as EUV for instance, checking the corresponding constraints may be cumbersome.\\

The core problem, when we ignore restrictions arising from any specific lithography technology (again we can introduce the corresponding constraints later on), is a new interesting extension of graph coloring. Given a graph $G=(V,E)$, $F\subseteq E$, and an integer $k\geq 1$, color the nodes of $G$ so that the connected component induced by each color admits a Hamiltonian path of length at most $k-1$. As $k$ grows, this seems to be a much more challenging problem to solve as it combines the difficulty of coloring with Hamiltonicity, as confirmed by our computational results (see Section \ref{experiments}). We again develop an integer model in the same vein as the previous one for $k\geq 2$.



\begin{align}
	&\min \sum_{i=1}^L \lambda^i && \label{general_01_obj}\\
	& \sum_{u\in N_F(v)}x^{i,\kappa}_{(v,u)} 										&& \leq \quad y^i_{v,v}, && \mbox{ if } \kappa=0, \forall i \in [L], \forall v \in V \label{general_01_cst1}\\
	& \sum_{u\in N_F(v)}x^{i,\kappa}_{(v,u)}-\sum_{u\in N_F(v)}x^{i,\kappa-1}_{(u,v)} && \leq \quad 0, && \forall \kappa: k-2 \geq \kappa \geq 1, \forall i \in [L], \forall v \in V \label{general_01_cst2}\\
	& \sum_{\kappa=0}^{k-2} (x^{i,\kappa}_{(u,v)}+ x^{i,\kappa}_{(v,u)}) && = \quad x^i_{(u,v)}, 	  && \forall i \in [L], \forall (u,v)\in F \label{general_01_cst3}\\
	& y^i_{v,v}+\sum_{\kappa=0,...,k-2} \sum_{u\in N_F(v)}x^{i,\kappa}_{(u,v)} && = \quad  z^i_v, 	  && \forall i \in [L], \forall v\in V \label{general_01_cst4}\\
	&   y^i_{u,o} + x^i_{(u,v)} -1											   && \leq \quad y^i_{v,o}, && \forall i \in [L], \forall (u,v)\in F, \forall o\in V \label{general_01_cst5}\\
	&   y^i_{v,o} + x^i_{(u,v)} -1 												&& \leq \quad y^i_{u,o}, && \forall i \in [L], \forall (u,v)\in F, \forall o\in V \label{general_01_cst6}\\
	&  \sum_{o \in V} y^i_{v,o} && =  z^i_v, && \forall i \in [L], v \in V \label{general_01_cst7}\\
	& y^i_{u,o} + \sum_{o' \in V, o'\neq o} y^i_{v,o'} && \leq \quad 1, && \forall i \in [L], \forall (u,v)\in E, \forall o\in V \label{general_01_cst8}\\
	& \sum_{i=1}^L z^i_v && =  1, && \forall v \in V \label{general_01_cst9}\\
	& x^{i,\kappa}_{(u,v)}, y^i_{u,o}, x^i_{(u,v)}, z^i_v && \leq \quad \lambda^i, && \forall i \in [L], \forall \kappa=0,...,k-2, \forall (u,v)\in F,  \forall o,u,v\in V \label{general_01_cst10}\\
	& x^{i,\kappa}_{(u,v)}, y^i_{u,o}, x^i_{(u,v)}, z^i_v, \lambda^i && \in \{0,1\}  && \forall i \in [L], \forall \kappa=0,...,k-2, \forall (u,v)\in F,  \forall o,u,v\in V
\end{align}

Variable $x^i_{(u,v)} \in \{0,1\}$  indicates whether an edge $(u,v) \in F$ is chosen in a path with color $i$. $x^{i,\kappa}_{(u,v)} \in \{0,1\}$ indicates whether an edge $(u,v) \in F$ is used as the $(\kappa+1)$-th edge in the direction from $u$ to $v$ in one of the disjoint paths of color $i$ (this again explicitly gives an orientation to the path). $y^i_{v,o} \in \{0,1\}$  indicates whether a node $v \in V$ is in a path starting in $o \in V$ with color $i \in [L]$. $y^i_{v,v} \in \{0,1\}$  indicates whether a node $v \in V$ is the first node of a path starting in $v \in V$ with color $i \in [L]$. $z^i_v \in \{0,1\}$ indicates whether a node $v \in V$ has color $i$. $\lambda^i \in \{0,1\}$ indicates whether a color $i$ is chosen.  

Constraints (\ref{general_01_cst1}), (\ref{general_01_cst2}), (\ref{general_01_cst3}), and (\ref{general_01_cst4}) have the same meaning as in the previous model. Constraints (\ref{general_01_cst5}) and (\ref{general_01_cst6}) identify the path each node belongs to by propagating connectivity, e.g., if $u$ is in a path of color $i$ starting in $o$ and $(u,v)$ is taken in color $i$ (in the direction from $u$ to $v$), then $v$ is in the path of color $i$ starting in $o$. Constraint (\ref{general_01_cst7}) imposes consistency for the value taken by $y^i_{v,o},$ (a vertex is in a path in color $i$ if and only if it is in color $i$). Constraint (\ref{general_01_cst8}) ensures that two adjacent nodes cannot belong to different paths of the same color. Finally, constraint (\ref{general_01_cst9}) imposes that each node $v \in V$ receives exactly one color, and constraint (\ref{general_01_cst10}) imposes that vertices and paths can be assigned to a color only if the color is selected. The number of variables in the formulation is in the order of $O(L.k.n^2)$, and the number of constraints is in the order of $O(L.n^3)$, where $n$ is the number of nodes of the (original) graph, and $k$ is the maximum number of nodes in the paths. 

\section{Numerical experiments}\label{experiments}

In this section, we report on our computational experiments with the various models described in the previous sections. We tested our models on ten instances, named clip1, \dots, clip10 arising from true industrial layouts at Mentor Graphics (the corresponding layouts are available upon request). The number of vias for each instance are given in Table \ref{numbers}.

\begin{table}[h!]
\scalebox{0.9}{\begin{tabular}{|l|l|l|l|l|l|l|l|l|l|l|}
\hline
Name & clip1 & clip2 & clip3 &clip4 &clip5 &clip6 &clip7 &clip8 &clip9 &clip10 \\
\hline
Number of vias & 200850& 173741& 236979& 184486& 214697& 248375& 225795& 126029& 200850& 173741\\
\hline
\end{tabular}}
\caption{Number of vias for each instance.}\label{numbers}
\end{table}

We do not use the true values for $Litho_{dist}$, $L_0$, and $U_0$ for confidentiality reasons. Instead, following \cite{Kuang,Yang}, we use three different values of $Litho_{dist}$ (31nm, 41nm, 49nm – note that distance is border to border), and set $L_0=20nm$ and $U_0=40nm$ (note the distance is center to center here). We re-scaled first the layout so that the minimum distance - border to border - between any two vias corresponds to a targeted pitch size of 10nm and so that the diameter of the vias is also 10nm, as in \cite{Kuang} (we assume here for simplicity that vias are disks, hence the distances border to border and center to center are easily obtained). We focus on 193 immersion and thus only consider DSA edges that are parallel to the x or y axis. We mainly focus on groups of size two and three, as this is a current limit imposed by manufacturing constraints, but we also discuss results for groups of size five to further evaluate the evolution of the performance of the models when $k$ increases (these results are more prospective). We set additional restrictions to our models to remove `L-shaped' guiding patterns (again to be compliant with the corresponding lithography technology that imposes guiding patterns to be parallel to the x and y axis), that is, for all triplets of vias $\{u,v,w\}$ such that $(u,v)$ and $(v,w)$ are DSA edges but where the angle formed by $u,v,w$ is 90 degree, adding the constraint $z^i_u + z^i_v+z^i_w \leq 2$ to our models (except for the model from Section \ref{small} where we simply removed the corresponding paths from the list).

Our models were implemented in CPLEX 12.6.3. In the first part of the study, we used the default parameters in CPLEX. We used the Networkx Python library to enumerate all possible paths up to a certain length for the model in Section \ref{small}. The tests were conducted on a machine equipped with an Intel(R) Xeon(R) CPU E5-2640 2.60 GHz and a memory of 529GB. We enforced a time limit of 3600 seconds (one of the models could solve all instances within this time limit, moreover, we ran the recalcitrant models for longer periods of time on the harder instances, and the results were similar). 

The main characteristics of the instances are described in Tables \ref{tab:31}, \ref{tab:41}, and \ref{tab:49}. 

\begin{table}[h!]
$$\begin{array}{l|l|l|l|l}
name &	 |V|	&|E|&  |F|		&  |E|/|V| \\
\hline
clip1&	200850&	198199&	198199&		0,99\\
clip2&	173741&	166867&	166865&		0,96\\
clip3&	236979&	215657&	206185&		0,91\\
clip4&	184486&	154235&	147546&		0,84\\
clip5&	214697&	208971&	208970&		0,97\\
clip6&	248375&	248323&	248323&		1\\
clip7&	225795&	216105&	194251&		0,96\\
clip8&	126029&	112741&	111695&		0,89\\
clip9&	200850&	198199&	198199&		0,97\\
clip10&	173741&	166867&	166865&		0,96\\
\end{array}$$
\caption{The number of nodes $|V|$, edges $|E|$, and DSA edges $|F|$ in the ten industrial instances when $Litho_{dist}$ = 31 nm}\label{tab:31}
\end{table}

\begin{table}[h!]
$$\begin{array}{l|l|l|l|l}
name &	 |V|	&|E|&  |F|		&  |E|/|V| \\
\hline
clip1&	200850&	389723&	198199&	1,94\\
clip2&	173741&	325605&	166867&	1,87\\
clip3&	236979&	370096&	215657&	1,56\\
clip4&	184486&	267536&	154235&	1,45\\
clip5&	214697&	430385&	208971&	2\\
clip6&	248375&	512810&	248323&	2,1\\
clip7&	225795&	371507&	216105&	1,65\\
clip8&	126029&	206612&	112741&	1,64\\
clip9&	200850&	389723&	198199&	1,94\\
clip10&	173741&	325605&	166867&	1,87\\
\end{array}$$
\caption{The number of nodes $|V|$, edges $|E|$, and DSA edges $|F|$ in the ten industrial instances when $Litho_{dist}$ = 41 nm}\label{tab:41}
\end{table}

\begin{table}[h!]
$$\begin{array}{l|l|l|l|l}
name &	 |V|	&|E|&  |F|		&  |E|/|V| \\
\hline
clip1&	200850&	514342&	198199&	2,56\\
clip2&	173741&	430739&	166867&	2,47\\
clip3&	236979&	489064&	215657&	2,06\\
clip4&	184486&	352820&	154235&	1,91\\
clip5&	214697&	521711&	208971&	2,43\\
clip6&	248375&	619021&	248323&	2,49\\
clip7&	225795&	530039&	216105&	2,35\\
clip8&	126029&	259120&	112741&	2,06\\
clip9&	200850&	514342&	198199&	2,56\\
clip10&	173741&	430739&	166867&	2,48\\
\end{array}$$
\caption{The number of nodes $|V|$, edges $|E|$, and DSA edges $|F|$ in the ten industrial instances when $Litho_{dist}$ = 49 nm}\label{tab:49}
\end{table}

The density of the graph is reported as $|E|/|V|$, as in \cite{Kuang,Yang}. Our industrial layouts actually exhibit a much larger density than the pseudo-industrial instances used in \cite{Kuang,Yang}. This makes a huge difference from a computational point of view. Indeed, the computational time is somewhat dominated by the largest connected component (obviously we can parallelize the computation to solve the problem on each connected component independently). The size of the largest connected component for each instance is reported in Tables \ref{tab:cc31}, \ref{tab:cc41}, and \ref{tab:cc49} (clip1\_31 represents the largest connected component of clip1 when $Litho_{dist}=31nm$, and so on). We do not have the figures for the instances used in \cite{Kuang,Yang}, however, what they consider as dense graphs are sparser than the sparsest graphs we consider here. This tends to indicate that the size of the largest and the average connected components in their benchmark are typically small, which explains why they have a computational time in the order of a few seconds for the overall instance without parallelization. In the following, we only document the characteristics and report the computational time on the largest connected component, as we believe this provides a better measure of the problem complexity. We also report the maximum clique size ($\omega$) and maximum degree ($\Delta$) of the corresponding instances.  

\begin{table}[h!]
$$\begin{array}{l|l|l|l|l|l|l}
 Instance &	|V| 	&|E| 	&|F|  &|E|/|F|	&\omega	&\Delta\\
 \hline
	clip1\_31&	191& 	242&		242	& 	1,27& 	3& 	5\\
	clip2\_31&	139&	 	188&		188	& 	1,35& 	3& 	5\\
	clip3\_31&	98&		117&		108	& 	1,19& 	3& 	4\\
	clip4\_31&	120&		147&		139	& 	1,22& 	3& 	4\\
	clip5\_31&	170&		213& 	213	& 	1,25& 	3& 	5\\
	clip6\_31&	178&		229& 	229	& 	1,29& 	3& 	5\\
	clip7\_31&	203&		256& 	223	& 	1,26& 	3& 	5\\
	clip8\_31&	122&		162& 	160	& 	1,33& 	3& 	5\\
	clip9\_31&	152&		193& 	193	& 	1,27& 	3& 	4\\
	clip10\_31&	139&		175& 	175	& 	1,26& 	4& 	4\\
\end{array}$$
\caption{Largest connected components in each instance for $Litho_{dist}$=31 nm}\label{tab:cc31}
\end{table}

\begin{table}[h!]
$$\begin{array}{l|l|l|l|l|l|l}
 Instance &	|V| 	&|E| 	&|F|  &|E|/|F|	&\omega	&\Delta\\
 \hline
clip1\_41&	1714&	3595&	1741&		2,1&		6&	9\\ 
clip2\_41&	1696&	3808&	1858&		2,25&	6&	9\\ 
clip3\_41&	1099&	1842&	947&			1,68&	4&	7\\ 
clip4\_41&	816&		1457&	787&			1,79&	4&	7\\ 
clip5\_41&	3850&	8641&	4030&		2,24&	6&	10\\ 
clip6\_41&	3598&	7751&	3587&		2,15&	6&	9\\ 
clip7\_41&	2337&	4507&	2238&		1,93&	4&	7\\ 
clip8\_41&	1033&	1899&	990&			1,84&	4&	7\\ 
clip9\_41&	1713&	3880&	1879&		2,27&	6&	9\\ 
clip10\_41&	1221&	2690&	1332&		2,2&		6&	9\\ 
\end{array}$$
\caption{Largest connected components in each instance for $Litho_{dist}$=41 nm}\label{tab:cc41}
\end{table}

\begin{table}[h!]
$$\begin{array}{l|l|l|l|l|l|l}
 Instance &	|V| 	&|E| 	&|F|  &|E|/|F|	&\omega	&\Delta\\
 \hline
clip1\_49&	15363&	40865&	15642&		2,659962247&	6&	11\\
clip2\_49&	14499&	38874&	14945&		2,681150424&	6&	10\\
clip3\_49&	3766&	8447&	3277&		2,242963356&	6&	9\\
clip4\_49&	1442&	3166&	1189&		2,19556172&	6&	9\\
clip5\_49&	16401&	40818&	16442&		2,488750686&	6&	11\\
clip6\_49&	18808&	48279&	19245&		2,5669396&	6&	11\\
clip7\_49&	16387&	39375&	14491&		2,402819308&	6&	11\\
clip8\_49&	3809&	8659&	3621&		2,273300079&	6&	9\\
clip9\_49&	14915&	39146&	14945&		2,624606101&	6&	11\\
clip10\_49&14473&	37609&	14522&		2,598562841&	6&	10\\
\end{array}$$
\caption{Largest connected components in each instance for $Litho_{dist}$=49 nm}\label{tab:cc49}
\end{table}

\subsection{k=2}

In this subsection, we compare our models assuming we can only create groups of at most size two. In this case, we can compare the models from Section \ref{pairing} and Section \ref{small}. We call {\em pairing} the model from Section \ref{pairing} and {\em na\"ive} the model from Section \ref{small}. We also provide the computational time for proper coloring. `Best value' indicates the best coloring found, `time to best' the time (in seconds) to find the best solution, `time to certify' the time (in seconds) to certify that the solution is optimal (the time limit when no certificate of optimality is obtained), and `cplex gap' the percentage between the best value and the best lower bound. When not even a first feasible solution is found (either because of memory or cpu limits), we use a backslash sign ($\backslash$).

\begin{table}[h!]
\includegraphics[width=\textwidth]{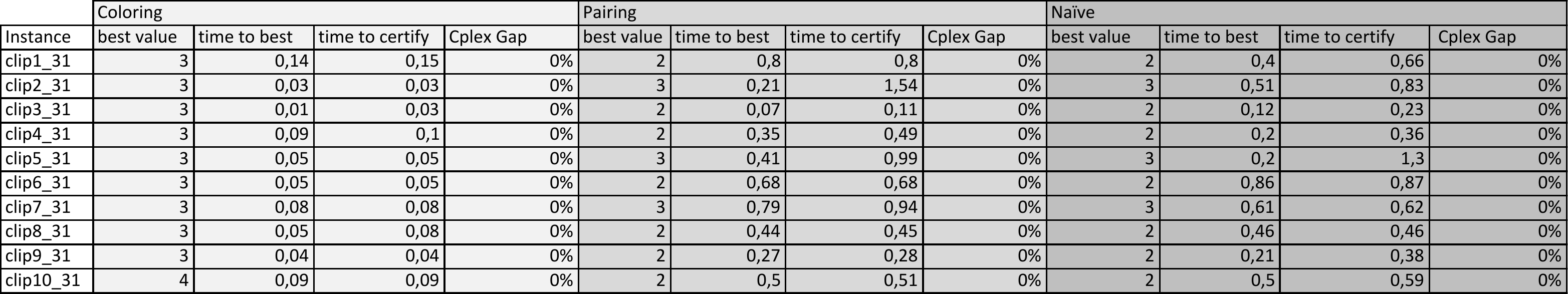}
\caption{Comparison of the pairing and the na\"ive model for $k=2$ and $Litho_{dist}$ = 31nm.}
\end{table}

\begin{table}[h!]
\includegraphics[width=\textwidth]{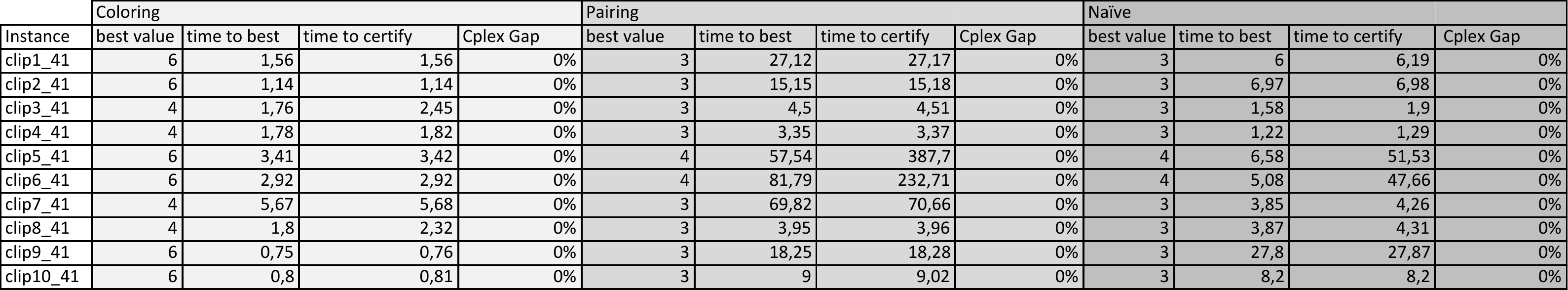}
\caption{Comparison of the pairing and the na\"ive model for $k=2$ and $Litho_{dist}$ = 41nm.}
\end{table}

\begin{table}[h!]
\includegraphics[width=\textwidth]{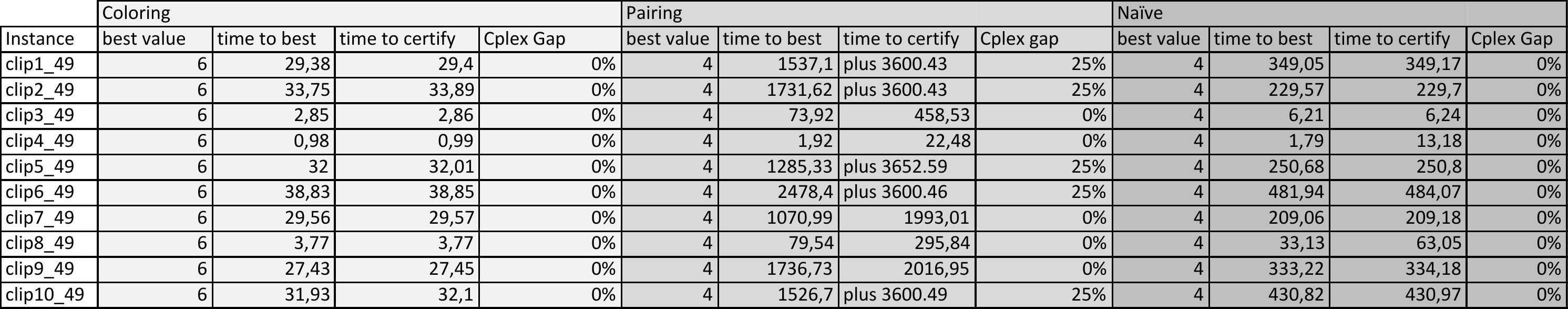}
\caption{Comparison of the pairing and the na\"ive model for $k=2$ and $Litho_{dist}$ = 49nm.}
\end{table}

We can observe from our experiments that the na\"ive model outperforms the pairing model. Indeed, while the computational times are similar on the sparsest instances ($Litho_{dist}$=31 nm), the difference becomes more obvious as the density increases. It was quite surprising at first to see that the pairing model cannot solve half of the largest instances within the time limit of 1 hour, while the na\"ive model can solve all instances within a few minutes (8 minutes at most). However, this is not completely unexpected as the na\"ive model allows convexifying the integer hull of the paths (this is also true for longer paths) at a price that is not too high when $k$ is small. Furthermore, we believe that CPLEX can better exploit the structure of the na\"ive model, as it combines the set partitioning and set packing constraints, both of which are well-studied, and many strong cuts for these problems are embedded in the CPLEX default settings. Interestingly, this analysis is also likely to explain the good performance of the na\"ive model when $k=3$, as we will see in the next section.
 
\subsection{k=3}

In this subsection, we compare our models assuming we can only create groups of at most size three. We focus on paths. In this case, we can compare the models from Section \ref{small}, Section \ref{large}, and Section \ref{general}. To eliminate any confusion, we call {\em na\"ive induced} the na\"ive model instantiated by listing all induced paths of at most length two (i.e., groups of at most length three), {\em na\"ive general} the na\"ive model instantiated by listing all groups of at most length three that exhibit a Hamiltonian path of length two, {\em induced} the model from Section \ref{large}, and {\em general} the model from Section \ref{general}.

\begin{table}[h!]
\includegraphics[width=\textwidth]{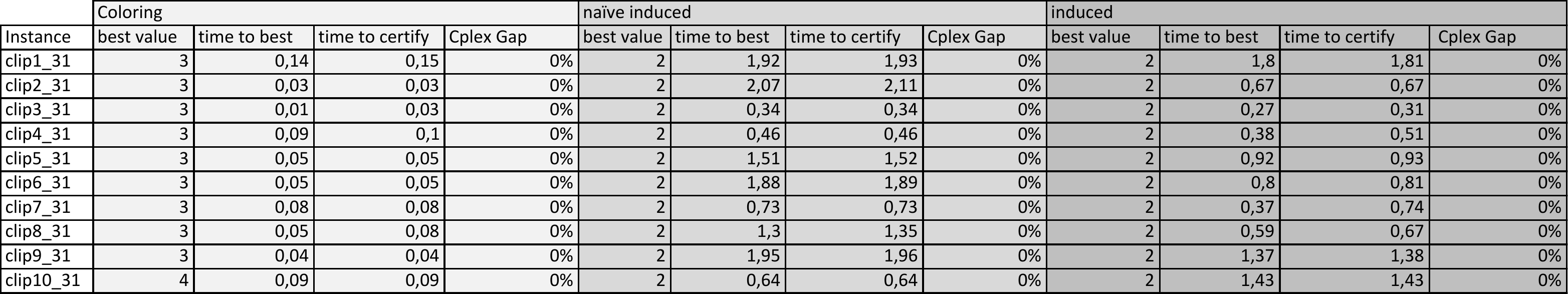}
\caption{Comparison of the na\"ive induced model and the induced model for $k=3$ and $Litho_{dist}$ = 31nm.}
\end{table}

\begin{table}[h!]
\includegraphics[width=\textwidth]{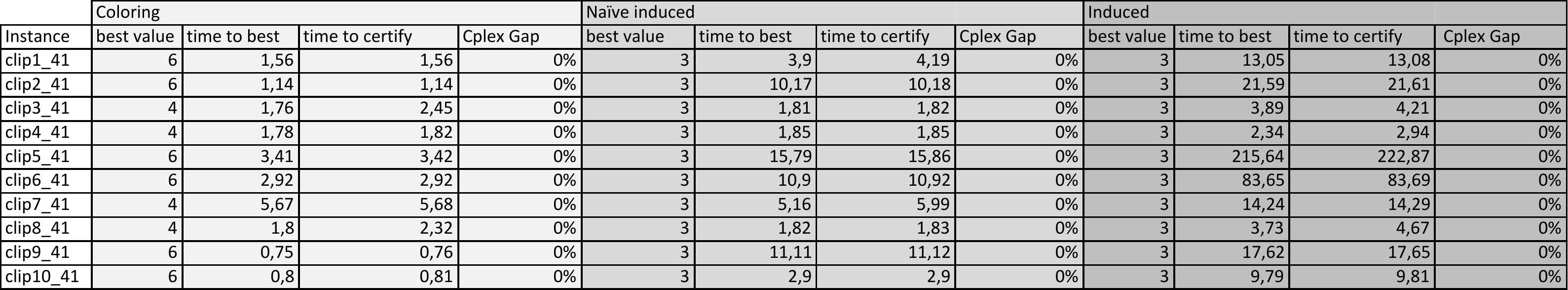}
\caption{Comparison of the na\"ive induced model and the induced model for $k=3$ and $Litho_{dist}$ = 41nm.}
\end{table}

\begin{table}[h!]
\includegraphics[width=\textwidth]{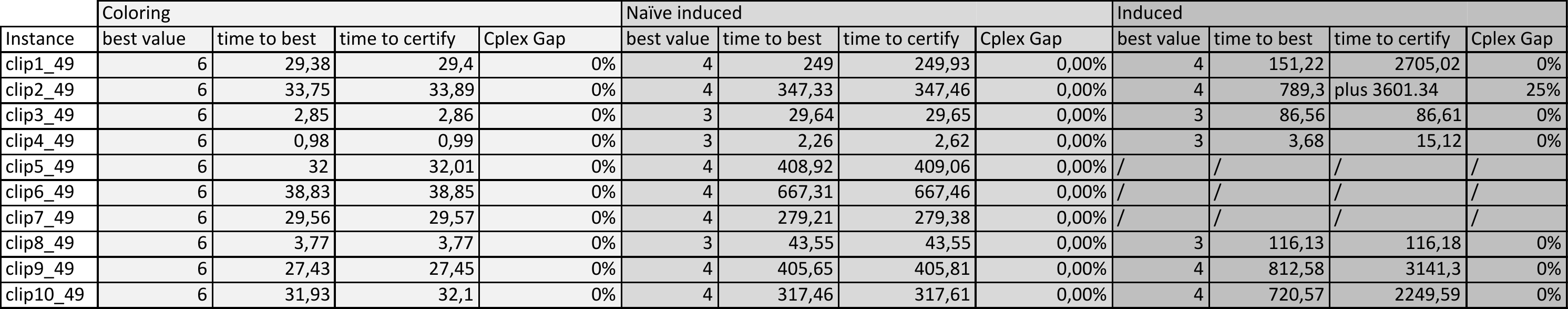}
\caption{Comparison of the na\"ive induced model and the induced model for $k=3$ and $Litho_{dist}$ = 49nm.}
\end{table}

We first focus on the case where the paths are induced and then move to the case where we allow for non-induced paths. Here again, the na\"ive induced model clearly outperforms the ad-hoc induced version. The results are somewhat surprising at first sight since while we would not expect the induced model to perform better than the na\"ive model on small groups, we did not expect it to already reach its limits for $k=3$ when the graphs are large. This calls into question the interest in such model and the existence of better models to cope with larger values of $k$. Indeed, we tried the models with $k=5$ and observed very similar behavior (our instances are not that well-suited to testing larger values of $k$, as the number of feasible paths does not increase by much when going from $k=3$ to $k=5$ or larger, and hence the na\"ive model will always perform similarly while the induced model will run out of memory even faster). We believe that the clear advantage of the na\"ive model again derives from the fact that CPLEX can exploit the set packing and set partitioning nature of the problem, and the fact that the formulation convexifies the path of length two.  

In the case where we allow for non-induced paths, the results are even more in favor of the na\"ive model, as shown in the following tables. Observe that there is no difference on the optimal coloring whether we allow non-induced paths or not. As noted in the introduction, this has been anticipated by practitioners due to the structure of the industrial instances.

\begin{table}[h!]
\includegraphics[width=\textwidth]{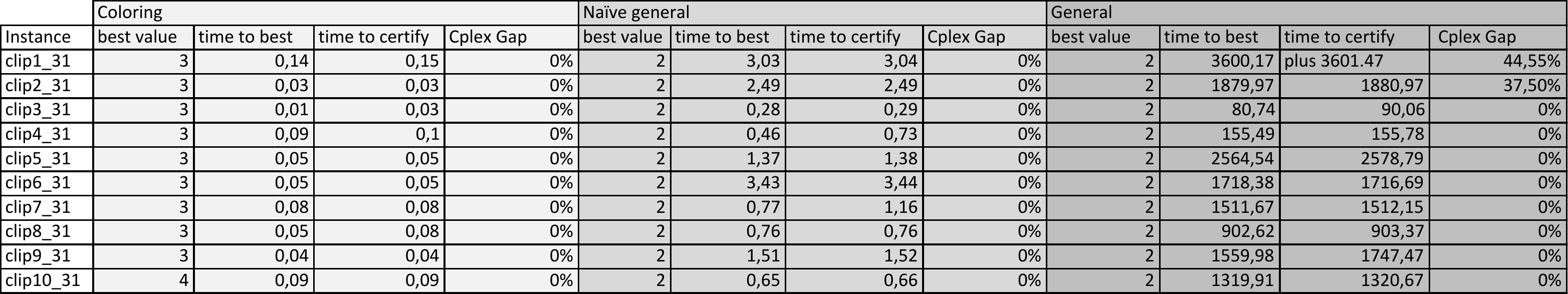}
\caption{Comparison of the na\"ive general model and the general model for $k=3$ and $Litho_{dist}$ = 31nm.}
\end{table}

\begin{table}[h!]
\includegraphics[width=\textwidth]{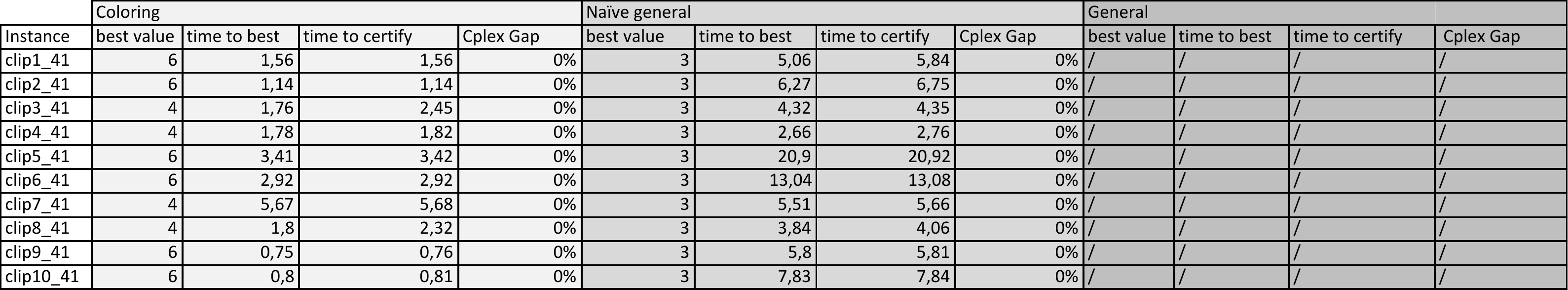}
\caption{Comparison of the na\"ive general model and the general model for $k=3$ and $Litho_{dist}$ = 41nm.}
\end{table}

\begin{table}[h!]
\includegraphics[width=\textwidth]{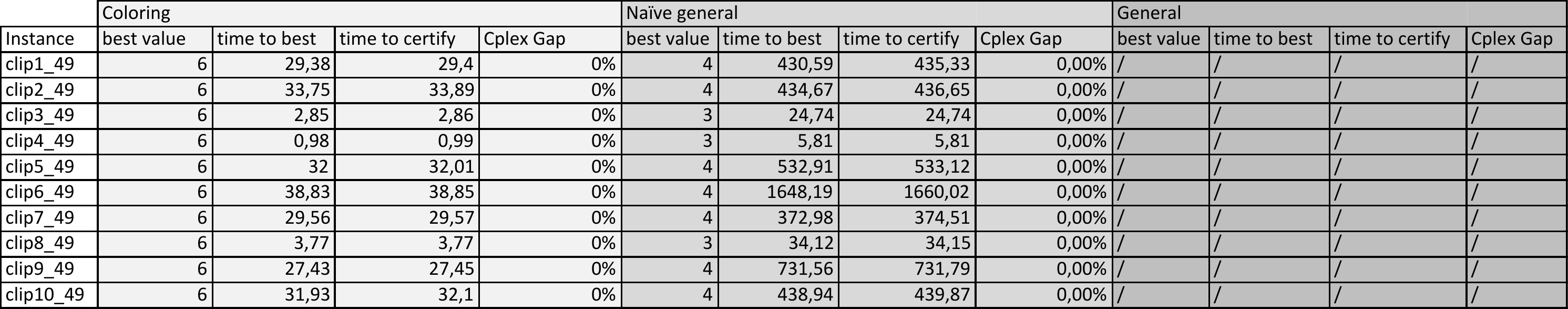}
\caption{Comparison of the na\"ive general model and the general model for $k=3$ and $Litho_{dist}$ = 49nm.}
\end{table}

Clearly, the general model seems highly inappropriate. Indeed, for large instances, it cannot even load in memory. This raises the question of the existence of more appropriate models, in the original space for instance, which could compete with the na\"ive model for small groups and would also allow dealing with larger groups. Relevant models are likely to  require exponentially many inequalities, and therefore, the use of cutting plane approaches. However, we leave the corresponding investigations for future research.

\subsection{Toward a column-generation approach for the na\"ive model}

One aspect that we have so far not really stressed is the fact that the na\"ive model relies on a complete enumeration of all feasible paths. Although we mentioned that we could enumerate the corresponding paths using the Networkx Python library, we have not commented on the time spent in this procedure, which, to be fair with respect to the other models, should be included in the computation time. In fact, including the pre-processing time (see Table \ref{prepro}) only marginally changes the conclusion. Nevertheless, it substantially increases the overall computation time.

\begin{table}[h!]
\centering \includegraphics[width=0.7\textwidth]{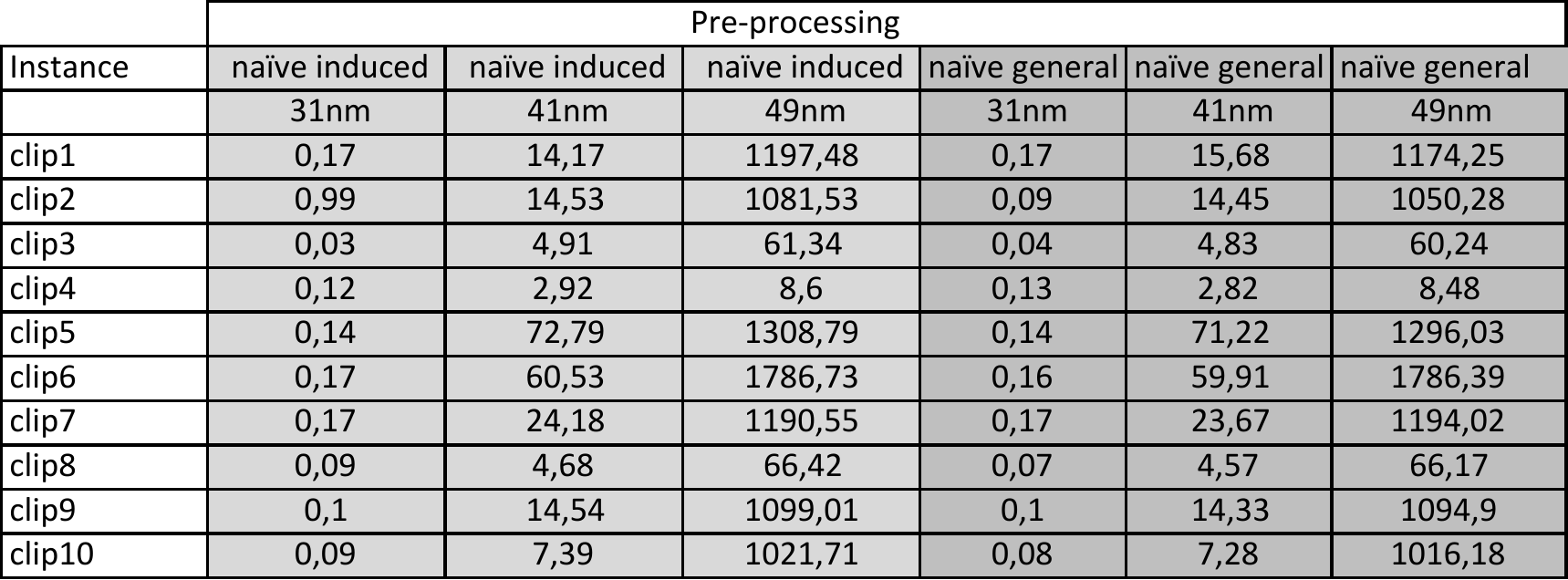}
\caption{Preprocessing times}\label{prepro}
\end{table}

Given the encouraging performance of the na\"ive model, we investigate it further. In particular, we evaluate the potential of applying a column generation approach to avoid listing all the feasible paths upfront. For such an approach to be successful, we need to evaluate the quality of the linear relaxation. In so doing, some clique inequalities are easily identifiable for this model, and the model in Section \ref{small} can easily be strengthened as follows.

\begin{align}
\min & \sum_{i=1}^L \lambda^i  \\
	&\sum_{i=1}^L \sum_{g \in \mathcal V: v\in g} x^i_u 		&& = \quad 1,	&& \mbox{for each via $v$} \label{naive2_cst1}\\
	&\sum_{f\in \mathcal V :u\in f} x^i_{f} + \sum_{g\in \mathcal V :v\in g} x^i_{g} 								&&\leq \quad 1, && \forall i \in [L], \forall  \mbox{u,v: $d(u,v)\leq L$} \label{naive2_cst2}\\
	&x^i_g 												&&\leq \quad \lambda^i, 	&& \forall i \in [L], \forall g\in {\mathcal V} \label{naive2_cst3}\\
	&x^i_g, \lambda^i 									&& \in  \quad \{0,1\}, 		&& \forall i \in [L], \forall g\in {\mathcal V}\\
\end{align}

Variable $\lambda^i \in \{0,1\}$ indicates whether color $i$ is chosen, and $x^i_g \in \{0,1\}$ indicates whether a group $g \in \mathcal{V}$ is colored with color $i$.
Constraint (\ref{naive2_cst1}) imposes that each via is assigned to exactly one group and one color. Constraint (\ref{naive2_cst2}) imposes that groups in conflict (for which there are two vias that are too close) have to receive different colors.  

We compared the performance of this new na\"ive model when all cuts are deactivated in CPLEX with the original model, with the default cuts activated. We now report the results for the general case with $k=3$ but a similar behavior is observed for the induced case and when $k=2$. It would seem that the new model with no additional cuts performs even better than the original model (with cuts activated). Again this might seem strange at first sight, but can partly be explained by the fact that the clique constraints we identified are probably quite strong already.  This is encouraging, as it tends to indicate that the corresponding linear relaxation is strong, and thus that a column-generation approach building on the later formulation might perform rather well, without requiring listing all paths upfront but instead generating the paths `on the fly' by solving a pricing problem. The corresponding promising approach is far beyond the scope of the current study, and we hence leave it for future investigations.

\begin{table}[h!]
\includegraphics[width=\textwidth]{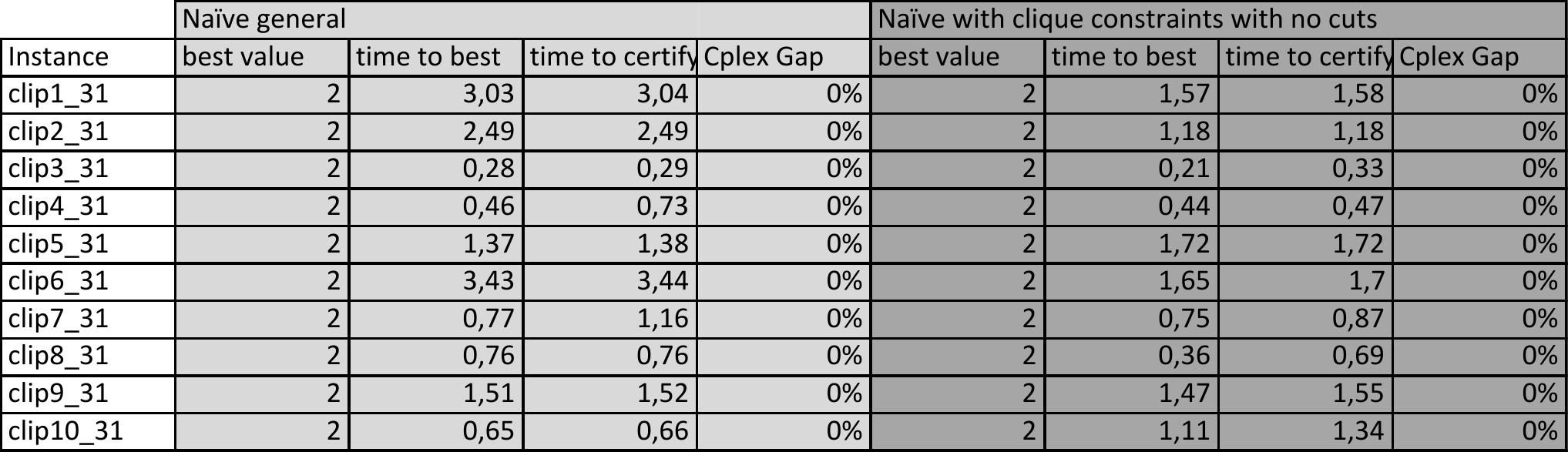}
\caption{Comparison of the original model with the new model with no CPLEX cuts for $k=3$ and $Litho_{dist}$ = 31nm.}
\end{table}

\begin{table}[h!]
\includegraphics[width=\textwidth]{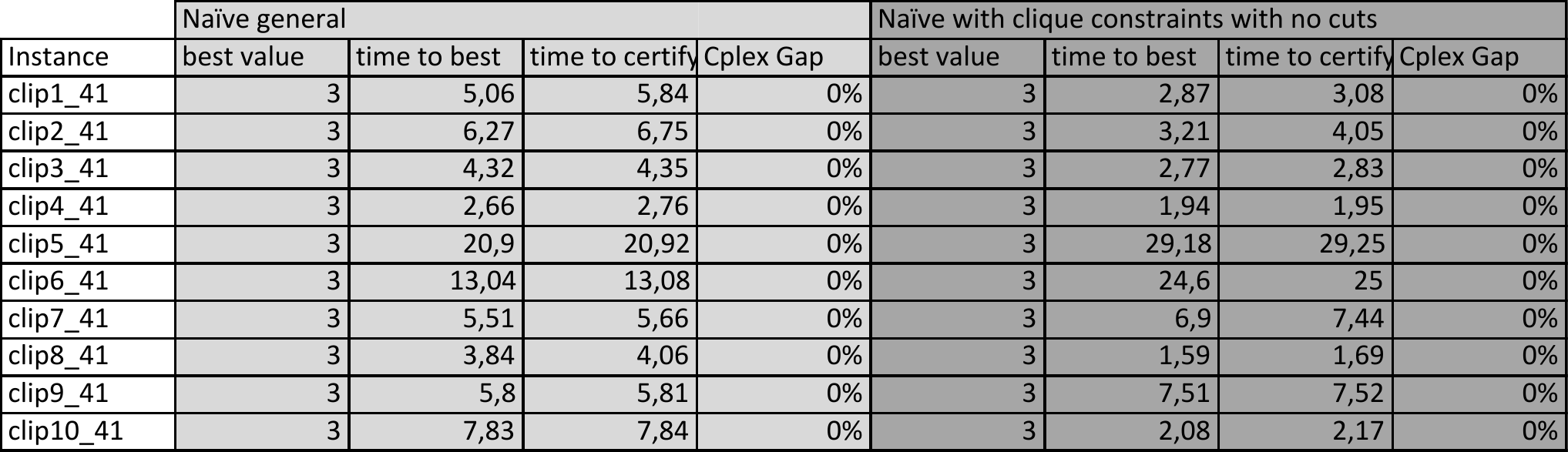}
\caption{Comparison of the original model with the new model with no CPLEX cuts for $k=3$ and $Litho_{dist}$ = 41nm.}
\end{table}

\begin{table}[h!]
\includegraphics[width=\textwidth]{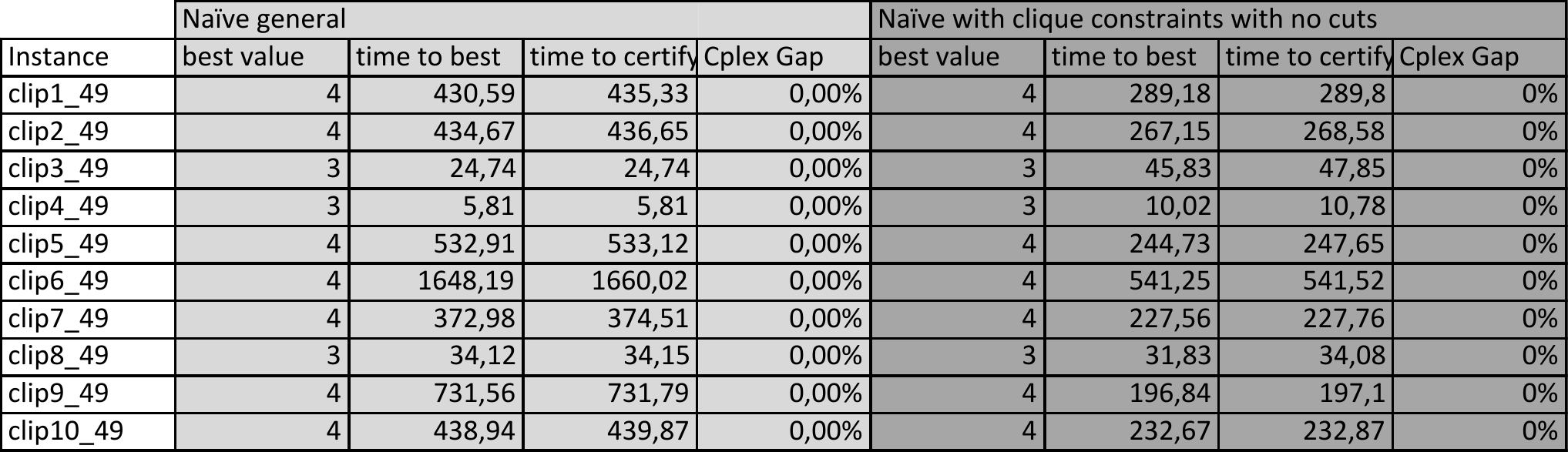}
\caption{Comparison of the original model with the new model with no CPLEX cuts for $k=3$ and $Litho_{dist}$ = 49nm.}
\end{table}

\section{Conclusion and perspectives}

In this study, we have developed several models for the manufacturing of vias through DSA-aware Multiple Patterning. Surprisingly, our computational experiments have shown that the most na\"ive models performed best on the industrial instances. Of course, this does not mean that other models should not be investigated further. Indeed, we only had access to a limited number of industrial cases that may not be representative of all possible instances. Furthermore, there might be other applications of our models beyond the manufacturing of vias. It would thus be interesting to develop new models that could scale better when $k$ increases. A possible line of research would be investigating models in the original space, which would certainly involve a large (exponential-size) number of constraints, and using such models in practice would therefore require cutting-plane approaches. Although developing the corresponding models and investigating their performance is beyond the scope of this paper, it would offer fertile ground for polyhedral studies. 

One of the main disadvantages of the na\"ive models is that they rely on a complete enumeration of the feasible groups upfront. In our applications, this was not problematic as the number of feasible paths was limited (due to restriction in size but also manufacturing constraints). Nevertheless, our investigations on the quality of the linear relaxation of these models suggest that a column-generation approach might be worth pursuing. In fact, not only should generating the path on the fly reduce the pre-processing time, but it should also considerably decrease the size of the models, which in turn could lead to substantial computational improvements. For the largest instances, the computation times were in the order of 8 minutes in the worst case. We believe that a column generation approach could bring the computational time down to a few tens of seconds, which would then be much more appealing from an industrial point of view (the tools could then be used in real time to evaluate different designs before the production process begins, for instance). 

In practice, while the computational time does not really allow using the corresponding models in production, Mentor Graphics used it to identify and improve weaknesses in their heuristics \cite{Dehia}. The heuristics that Mentor Graphics developed exploit the structure of the graphs arising from industrial applications. In this study, we have not attempted to exploit this line of research. Nevertheless, as the graphs are extremely sparse, there are many small cuts (isthmus, for instance) in connected components. Exploiting these structures by decomposing the problem further through Lagrangian relaxation and/or pure clustering ideas could significantly decrease the overall computation time by reducing the core problem to instances with only a few tens of vertices. This is certainly a direction worth investigating (an approach that has already been successfully investigated in \cite{Kuang} but on sparser instances). Moreover, there might be additional structures to explicitly exploit. We observed in many instances that the sparse graphs are `close to trees', and deem that the case of graphs with small tree-width is particularly relevant from an application perspective. We are currently investigating algorithms that exploit such property. In particular, we may prove that the $k$-path coloring problem is polynomial in this case and develop efficient dynamic programming algorithms \cite{Dehia, DP}. Other structures such as those identified in \cite{grid} may be worth investigating. 

A side benefit of the na\"ive model is that it allows introducing additional `validated' guiding patterns and easily adding other constraints. For instance, we have not considered constraints between groups in different masks. However, depending on the technology used, there might be additional constraints to take into account. For instance, two pairs of vias may lead to two guiding patterns that intersect, and this may be forbidden by the technology even if they belong to two different masks (see \cite{BadrPHD} for constraints of this type, called {\em mutually exclusive}). Although we have not considered such constraints thus far, they are easy to introduce in the na\"ive model (but more difficult in others).

Finally, from a practical perspective, there might be relevant alternative models for the manufacturing of vias. We mentioned the industry’s interest in minimizing the number of conflicts when fixing the number of patterning steps (such conflict could then possibly be removed by slightly adjusting the layout, for instance). Another interesting option would be to consider the problem of maximizing the minimum distance between any two features within a mask when the number of patterning steps is again fixed. This would allow identifying which lithography technology is more appropriate for the corresponding design, and if no technology is feasible, again identify small adjustments in the layout that may result in a feasible solution.  

\section{Acknowledgment}

We would like to thank Fran\c cois Clautiaux and St\'ephane Dauz\`ere-Peres for their very constructive comments and discussions on our work. This project has been partly supported by the Association Nationale de la Recherche et de la Technologie (Convention CIFRE 2015/0553).

\bibliographystyle{abbrv}
\bibliography{bib_graph,bib_semic}

\end{document}